\documentclass[10pt,journal,compsoc]{IEEEtran}
\usepackage[utf8]{inputenc}
\usepackage{balance}
\usepackage{graphicx}
\usepackage{amsmath}
\usepackage{amssymb}
\usepackage{setspace}
\usepackage{enumerate}
\usepackage{microtype}
\usepackage{hyperref}
\usepackage{cite}
\usepackage{color}
\usepackage{subcaption}
\usepackage[usenames,dvipsnames]{xcolor}
\usepackage{tcolorbox}
\usepackage{tabularx}
\usepackage{array}
\usepackage{colortbl}
\usepackage{multirow}
\newcolumntype{C}[1]{>{\centering\let\newline\\\arraybackslash\hspace{0pt}}m{#1}}

\DeclareMathOperator*{\argmin}{arg\,min}
\DeclareMathOperator*{\mean}{mean}

\tcbuselibrary{skins}

\begin{document}
\title{Adaptive Area-Preserving Parameterization of Open and Closed Anatomical Surfaces} 
\author{Gary P. T. Choi, Amita Giri, and Lalan Kumar\\
\thanks{This work was supported in part by the National Science Foundation under Grant No.~DMS-2002103 (to G. P. T. Choi), and the Prime Ministers Research Fellowship (PMRF), Government of India (to A. Giri).}
\thanks{G. P. T. Choi is with the Department of Mathematics, Massachusetts Institute of Technology, Cambridge, MA, USA (email: ptchoi@mit.edu).}
\thanks{A. Giri is with the Department of Electrical Engineering, Indian Institute of Technology Delhi, New Delhi, India (email: Amita.Giri@ee.iitd.ac.in).}
\thanks{L. Kumar is with the Department of Electrical Engineering and Bharti School of Telecommunication, Indian Institute of Technology Delhi, New Delhi, India (email:lkumar@ee.iitd.ac.in).}
}

\IEEEtitleabstractindextext{%
\begin{abstract}
The parameterization of open and closed anatomical surfaces is of fundamental importance in many biomedical applications. Spherical harmonics, a set of basis functions defined on the unit sphere, are widely used for anatomical shape description. However, establishing a one-to-one correspondence between the object surface and the entire unit sphere may induce a large geometric distortion in case the shape of the surface is too different from a perfect sphere. In this work, we propose adaptive area-preserving parameterization methods for simply-connected open and closed surfaces with the target of the parameterization being a spherical cap. Our methods optimize the shape of the parameter domain along with the mapping from the object surface to the parameter domain. The object surface will be globally mapped to an optimal spherical cap region of the unit sphere in an area-preserving manner while also exhibiting low conformal distortion. We further develop a set of spherical harmonics-like basis functions defined over the adaptive spherical cap domain, which we call the adaptive harmonics. Experimental results show that the proposed parameterization methods outperform the existing methods for both open and closed anatomical surfaces in terms of area and angle distortion. Surface description of the object surfaces can be effectively achieved using a novel combination of the adaptive parameterization and the adaptive harmonics. Our work provides a novel way of mapping anatomical surfaces with improved accuracy and greater flexibility. More broadly, the idea of using an adaptive parameter domain allows easy handling of a wide range of biomedical shapes.
\end{abstract}

}

\maketitle

\IEEEdisplaynontitleabstractindextext

\IEEEraisesectionheading{\section{Introduction}\label{sec:num1}}
\IEEEPARstart{S}{urface} parameterization is the process of mapping a complicated surface to a simple parameter domain, which plays an important role in biomedical visualization~\cite{angenent1999laplace,halier2000nondistorting,kreiser2018survey} and shape morphometry~\cite{styner2006framework,chung2008tensor,choi2020tooth,choi2020shape}. In many situations, the parameterization is desired to be with low geometric distortion. However, by a classical result of differential geometry~\cite{do2016differential}, it is in general impossible to achieve an isometric (distance preserving) parameterization. We can only achieve an angle-preserving (conformal) map, an area-preserving (authalic) map, or a balance between area and angle preservation. Over the past several decades, numerous parameterization algorithms have been developed~\cite{floater2005surface,sheffer2007mesh}. In particular, there has been a vast number of works on conformal parameterization algorithms for mapping genus-0 closed surfaces onto the unit sphere~\cite{angenent1999conformal,haker2000conformal,gu2004genus,lai2014folding,choi2015flash,choi2016spherical,choi2016fast,choi2020parallelizable} and simply-connected open surfaces onto a planar domain such as the unit disk~\cite{choi2015fast,yueh2017efficient,choi2018linear}, a rectangle~\cite{meng2016tempo}, a prescribed non-convex template~\cite{choi2017conformal}, or a domain with minimal area distortion~\cite{sawhney2017boundary}. However, while conformal mappings preserve angles and hence the local geometry of surfaces, the area distortion they produce may be highly undesirable. For instance, highly squeezed regions under a conformal parameterization may lead to inaccuracies in the surface harmonics representations. Therefore, some recent works have focused on the computation of area-preserving parameterizations for genus-0 closed surfaces~\cite{zhao2013area,cui2019spherical,pumarola20193dpeople} and simply-connected open surfaces~\cite{zou2011authalic,choi2018density,yueh2019novel}. Furthermore, area-preserving parameterizations have been found useful for biomedical visualization~\cite{zhu2003area,su2013area,choi2020area} as particular regions of biomedical structures will less likely to be shrunk under area-preserving mappings. More recently, a few works have considered the parameterization of biomedical surfaces onto other target domains. For instance, Nadeem \emph{et al.} developed a method called LMap~\cite{nadeem2017lmap} that flattens a local selected region-of-interest instead of the entire surface. Also, Giri \emph{et al.} proposed two area-preserving parameterization methods for open and closed anatomical surfaces with the target parameter domain being a hemisphere~\cite{giri2021open}. However, the above-mentioned parameterization methods only focus on reducing the geometric distortion with a target parameter domain determined \emph{a priori}. It is natural to ask whether one can parameterize the surface globally onto an \emph{adaptive} parameter domain, where the overall shape of the parameter domain is also a variable that we can optimize throughout the parameterization process.  

In this work, we propose two adaptive area-preserving parameterization methods for simply-connected open and closed anatomical surfaces. More specifically, we consider parameterizing any given simply-connected surface onto an adaptive spherical cap $\mathbb{S}^2_{Z \geq Z^*} = \{(X,Y,Z) \in \mathbb{R}^3: X^2+Y^2+Z^2 = 1 \text { and } Z \geq Z^*\}$ in an area-preserving manner, where the lower bound $Z^*$ is automatically determined. We also develop a set of spherical harmonics (SH)-like basis functions defined over the adaptive spherical cap domain exactly, which we call the \emph{adaptive harmonics} (AH). The novel combination of the adaptive surface parameterization and AH enables efficient anatomical shape description and reconstruction. When compared to the existing parameterization methods with fixed target shape, our methods are more flexible as the extra degree of freedom in the shape of the spherical cap allows us to further reduce the geometric distortion of the parameterization. Also, when compared to the existing parameterization methods with optimized target shape, our adaptive spherical cap domains are more standardized and hence are easier to utilize for various applications. Overall, the proposed parameterization methods achieve an optimal balance between flexibility and canonicity.

The rest of the paper is organized as follows. In Section~\ref{sect:background}, we introduce the theory of conformal geometry, quasi-conformal geometry and optimal mass transport. In Section~\ref{sect:main}, we describe our proposed framework for the adaptive area-preserving parameterization of open and closed surfaces and the formulation of AH. Experimental results on various anatomical surfaces are presented in Section~\ref{sect:experiment}. In Section~\ref{sect:conclusion}, we conclude the paper and discuss possible future directions.

\begin{figure*}[t!]
    \centering
    \includegraphics[width=\textwidth]{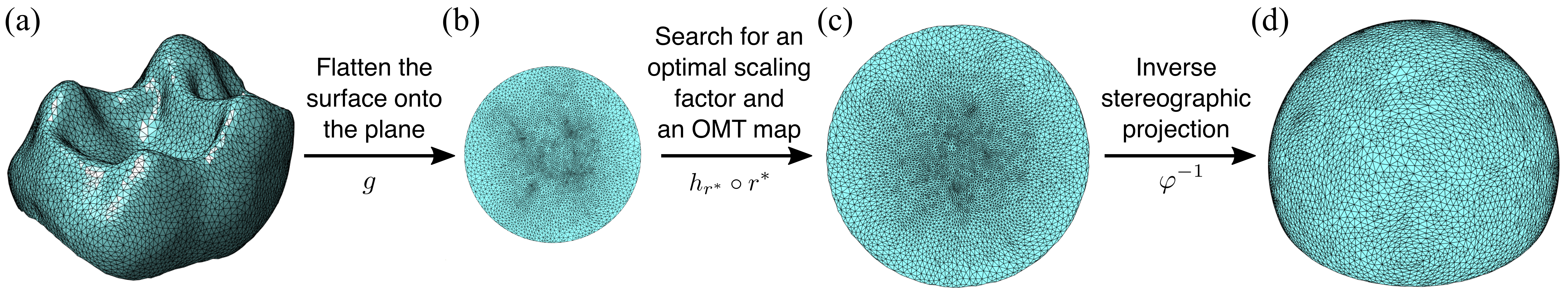}
    \caption{An illustration of the proposed adaptive area-preserving parameterization methods for open and closed anatomical surfaces. Given any simply-connected anatomical surface (see (a)), we first compute an initial flattening map onto the plane (see (b)). In case the surface is a closed surface, an extra step of puncturing a quadrilateral region~\cite{giri2021open} is applied beforehand. Next, we search for an optimal scaling factor and an OMT map simultaneously by solving an optimization problem (see (c)). Finally, we apply the inverse stereographic projection to obtain an adaptive area-preserving parameterization onto a spherical cap region (see (d)). The tooth surface shown here is adopted from MorphoSource~\cite{winchester2014dental,gao2015hypoelliptic}.}
    \label{fig:illustration}
\end{figure*}

\section{Background} \label{sect:background}
\subsection{Conformal and Quasi-Conformal Geometry}
In this section, we first introduce some important concepts in conformal and quasi-conformal geometry related to our work. Readers are referred to~\cite{lehto1973quasiconformal,ahlfors2006lectures} for details.

Mathematically, \emph{conformal} maps are mappings that locally preserve angles. Let $f:\mathbb{C} \to \mathbb{C}$ be a holomorphic function with $f(x,y) = u(x,y) + i v(x,y)$, where $u,v$ are real-valued functions and $i$ is the imaginary number with $i^2 = -1$. $f$ is conformal if it satisfies the Cauchy--Riemann equation
\begin{equation}
    \frac{\partial u}{\partial x} = \frac{\partial v}{\partial y} \ \ \text{ and } \ \ \frac{\partial v}{\partial x} = -\frac{\partial u}{\partial y}.
\end{equation}
More generally, conformal maps between two surfaces can be defined using their local charts. Two well-known examples of conformal maps are the stereographic projection and the inverse stereographic projection, which establish a one-to-one correspondence between the unit sphere and the extended complex plane. Denote the Cartesian coordinates of a point on the sphere and the corresponding point on the plane by $(X,Y,Z)$ and $(x,y)$ respectively. The stereographic projection $\varphi: \mathbb{S}^2 \to \overline{\mathbb{C}}$ is given by
\begin{equation} \label{eqt:stereographic}
    (x,y) = \varphi(X,Y,Z) = \left(\frac{X}{1-Z}, \frac{Y}{1-Z}\right),
\end{equation}
and the inverse stereographic projection $\varphi^{-1}: \overline{\mathbb{C}} \to \mathbb{S}^2$ is given by
\begin{equation} \label{eqt:stereographic_inverse}
\begin{split}
     (X,Y,Z) &= \varphi^{-1}(x,y) \\
     &= \left(\frac{2x}{1+x^2+y^2}, \frac{2y}{1+x^2+y^2}, \frac{-1+x^2+y^2}{1+x^2+y^2}\right).
\end{split}
\end{equation}

Quasi-conformal maps are a generalization of conformal maps. A map $f:\mathbb{C} \to \mathbb{C}$ is said to be \emph{quasi-conformal} if it satisfies the Beltrami equation
\begin{equation}
    \frac{\partial f}{\partial \overline{z}} = \mu(z) \frac{\partial f}{\partial z},
\end{equation}
where $\frac{\partial f}{\partial \overline{z}} = \frac{1}{2}\left(\frac{\partial f}{\partial x}  + i\frac{\partial f}{\partial y}\right)$, $\frac{\partial f}{\partial z} = \frac{1}{2}\left(\frac{\partial f}{\partial x}  - i\frac{\partial f}{\partial y}\right)$, and $\mu$ is a complex-valued function (called the \emph{Beltrami coefficient}) with $\|\mu\|_{\infty} < 1$. Here, $|\mu|$ captures the conformal distortion of $f$ in the sense that $|\mu| = 0$ if and only if $f$ is conformal. Analogous to conformal maps, quasi-conformal maps can be defined between surfaces with the aid of the local charts.

\subsection{Optimal Mass Transport and Area-Preserving Map}
The theory of optimal mass transport (OMT) has been studied for over two centuries~\cite{monge1781memoire,kantorovich1942translocation}, and recently it has been shown to be closely related to the computation of area-preserving mappings~\cite{zhao2013area,gu2016variational}. Let $\mathcal{X}$ and $\mathcal{Y}$ be two metric spaces with measures $\sigma, \tau$ respectively, and assume that $\mathcal{X}$ and $\mathcal{Y}$ have equal total measures, i.e. $\int_{\mathcal{X}} \sigma = \int_{\mathcal{Y}} \tau$. The transportation cost of moving $\mathbf{x}\in \mathcal{X}$ to $\mathbf{y} \in \mathcal{Y}$ is denoted by $c(\mathbf{x},\mathbf{y})$. A map $T:\mathcal{X} \to \mathcal{Y}$ is an \emph{optimal mass transport} map if it satisfies $\tau(B) = \sigma(T^{-1}(B))$ for all $B \subset \mathcal{Y}$ and minimizes the total transportation cost 
\begin{equation}\label{eqt:monge}
    \mathcal{C}(T) = \int_{\mathcal{X}} c(\mathbf{x},T(\mathbf{x})) \sigma(\mathbf{x}) d\mathbf{x}.
\end{equation}
By considering $\mathcal{X}$ as a surface and $\mathcal{Y}$ as a target parameter domain, the OMT map $T$ can be viewed as a parameterization mapping. In~\cite{kantorovich1948on}, Kantorovich introduced a relaxation of the Monge problem~\eqref{eqt:monge} and proved the existence and uniqueness of the OMT map. In~\cite{brenier1991polar}, Brenier showed that the OMT map is the gradient map of a convex function.

The discrete OMT mapping $T:\mathcal{X} \to \mathcal{Y}$ can be obtained using the approach in~\cite{zhao2013area}, which is based on the Monge--Brenier theory~\cite{brenier1991polar} and the variational principle in~\cite{gu2016variational}. More specifically, let $\mathbf{y}_1, \dots, \mathbf{y}_n \in \mathcal{Y}$ and $\tau$ be a discrete measure with delta masses at all $\mathbf{y}_i$, i.e. $\tau = \sum_{i=1}^n \tau_i \delta(\mathbf{y}-\mathbf{y}_i)$, and define the height vector $\textbf{h} = (h_1, \dots, h_n) \in \mathbb{R}^n$. Consider the energy $u_{\textbf{h}}(\mathbf{x}) = \max_{1 \leq i \leq n} (\langle \mathbf{x}, \mathbf{y}_i \rangle + h_i)$, where $\langle \cdot , \cdot \rangle$ denotes the inner product. It can be shown that $u_{\textbf{h}}$ is a convex function and is associated with a convex polyhedron with supporting hyperplanes given by $\langle \mathbf{x}, \mathbf{y}_i \rangle + h_i = 0$. Moreover, the energy 
\begin{equation}\label{eqt:omt_energy}
E(\textbf{h}) = \int_{\Omega}u_{\textbf{h}}(\mathbf{x}) \sigma (\mathbf{x}) d\mathbf{x} - \sum_{i=1}^n \tau_i h_i,
\end{equation}
where $\Omega = \text{supp } \sigma = \{\mathbf{x} \in \mathcal{X}: \sigma(\mathbf{x}) > 0\}$, is a convex energy. By minimizing $E$, the gradient map $\nabla u_{\textbf{h}}$ gives the desired OMT mapping. In practice, the computation of $\nabla u_{\textbf{h}}$ can be further simplified as the computation of the power diagram, i.e. the Voronoi diagram with the power distance $\text{Pow}(\mathbf{x},\mathbf{y}_i) = \frac{1}{2} \|\mathbf{x}-\mathbf{y}_i\|^2 - \frac{1}{2} h_i$. One can then use gradient descent to iteratively update $\textbf{h}$ and compute the power diagram until the energy $E(\textbf{h})$ is minimized. More details of the computational procedure can be found in~\cite{zhao2013area}.

A map $T:\mathcal{X} \to \mathcal{Y}$ is said to be an \emph{area-preserving} map if its Jacobian $J_T$ satisfies $|\det J_T| = 1$. By setting the source measure $\sigma$ and the target measure $\tau$ based on the local area of the surface and the target domain, one can obtain an area-preserving parameterization by solving an OMT problem.

\section{Proposed methods}\label{sect:main}
In this section, we describe our proposed methods for the adaptive area-preserving parameterization of open and closed surfaces, as well as the formulation of AH. The proposed parameterization methods are based on our recent work~\cite{giri2021open}, with a novel optimization step added for achieving the adaptive parameterization. The main features of our proposed methods are highlighted below:
\begin{enumerate}[(i)]
    \item Unlike other prior parameterization methods, the shape of the adaptive spherical cap parameter domain is automatically determined by our proposed methods.
    \item The parameterization of the open/closed surface onto the adaptive spherical cap domain is area-preserving.
    \item The parameterization also achieves a minimal conformal distortion.
    \item The parameterization can be naturally combined with AH for effective shape description.
\end{enumerate}
An illustration of the proposed adaptive area-preserving parameterization methods is given in Fig.~\ref{fig:illustration}. The detail of each step is provided in the following subsections.

\subsection{Adaptive Area-Preserving Parameterization of Simply-Connected Open Surfaces}
Let $\mathcal{S}_o$ be a simply-connected open surface (see Fig.~\ref{fig:illustration}(a)). The goal is to compute an area-preserving map of $\mathcal{S}_o$ onto an optimal spherical cap domain $\mathbb{S}^2_{Z \geq Z^*}$ with a hollow bottom part. 

\subsubsection{Initial flattening map}
The first step of the proposed algorithm is to flatten $\mathcal{S}_o$ onto a planar domain so as to simplify the subsequent computations. Since the boundary of an open spherical cap is a circle, it is natural to consider flattening $\mathcal{S}_o$ onto a planar disk domain as the initial map. Here, we use the disk conformal mapping method~\cite{choi2015fast} to compute a flattening map $g: \mathcal{S}_o \to \mathbb{D}$ onto the unit disk (see Fig.~\ref{fig:illustration}(b)). An advantage of the mapping method is that the mapping is bijective and conformal, and hence the resulting map is good enough for the next steps. Also, the computation is highly efficient. 

\subsubsection{Optimization on the plane}
Once we have obtained the initial flattening map $g$, we consider solving an optimization problem on the plane which yields an adaptive spherical cap parameterization with area preserved. This is achieved by searching for an optimal scaling factor for the planar disk domain and an OMT map onto it simultaneously. 

Here, our key observation is that by changing the radius $r$ of the disk, we can associate the disk with a unique spherical cap shape via the inverse stereographic projection $\varphi^{-1}$ by Eq.~\eqref{eqt:stereographic_inverse}. For any given $r$, we can always solve for an OMT mapping $h_r: r\mathbb{D} \to r\mathbb{D}$ from a disk with radius $r$ to itself. Similar to our recent work~\cite{giri2021open}, the source measure $\sigma$ and the target measure $\tau_r$ are carefully set in the computation of the OMT map as detailed below to ensure that the final parameterization $f$ is area-preserving. 

As for the source measure, since the planar domain will be projected onto a spherical cap by $\varphi^{-1}$, it is necessary to take the conformal factor of the inverse stereographic projection into consideration~\cite{nadeem2017spherical}. Hence, we set the source measure $\sigma$ to be
\begin{equation} \label{eqt:source_measure}
    \sigma = \frac{4 ~dx ~dy}{(1+x^2+y^2)^2},
\end{equation}
where $(x,y)$ are the Cartesian coordinates of the plane. As for the target measure, since an area-preserving map is desired, the target measure at every vertex $v_i$ should be set based on the local vertex area of it. It may be noted that since the final spherical cap is a subset of the unit sphere $\mathbb{S}^2$, in general, the total area of it is different from the total area of the input surface. Therefore, for any given radius $r$ of the disk, we set the target measure $\tau_r$ at every vertex as follows~\cite{nadeem2017spherical}:
\begin{equation} \label{eqt:target_measure_r}
    \tau_r(v_i) = \frac{\sum_{i = 1}^k 4r\mathcal{A}_{g(v_i)}/(1+r^2|g(v_i)|^2)}{\sum_{i = 1}^k \mathcal{A}_{v_i}} \mathcal{A}_{v_i}, 
\end{equation}
where $\mathcal{A}_{v_i}$ and $\mathcal{A}_{g(v_i)}$ are respectively the vertex area of $v_i$ and $g(v_i)$. More specifically, they are defined as the sum of the area of all triangles in the one-ring neighborhood divided by 3:
\begin{equation}
    \resizebox{0.95\hsize}{!}{$
    \displaystyle 
    \mathcal{A}_{v_i} = \frac{1}{3} \sum_{T \in \mathcal{N}(v_i)} \text{Area}(T), \ \mathcal{A}_{g(v_i)} = \frac{1}{3} \sum_{T' \in \mathcal{N}(g(v_i))} \text{Area}(T').
    $}
\end{equation}
Here the normalization factor $\frac{\sum_{i = 1}^k 4r\mathcal{A}_{g(v_i)}/(1+r^2|g(v_i)|^2)}{\sum_{i = 1}^k \mathcal{A}_{v_i}}$ corrects the overall area difference of the input surface and the spherical cap. We then follow the approach in~\cite{zhao2013area} and minimize the energy $E$ in Eq.~\eqref{eqt:omt_energy} with the source measure $\sigma$ and target measure $\tau_r$ in Eq.~\eqref{eqt:source_measure} and Eq.~\eqref{eqt:target_measure_r} to obtain the OMT mapping $h_r$.

Note that the above OMT mapping procedure works for any given disk of radius $r$. With this extra degree of freedom, we can now search for an optimal $r$ such that the corresponding area-preserving parameterization is the least geometrically distorted. Since a mapping is isometric if and only if it is both area-preserving and conformal, it is natural to consider using the conformal distortion as the criterion for the search of the optimal $r$. More specifically, we solve the following optimization problem:
\begin{equation}\label{eqt:optimize_r}
    r^* = \argmin_r \int |\mu_{(h_r \circ rg)^{-1}}(z)|^2 dz,
\end{equation}
where $\mu_{(h_r \circ rg)^{-1}}$ is the Beltrami coefficient of the mapping $(h_r \circ rg)^{-1}$. Recall that by quasi-conformal theory, $|\mu|$ captures the deviation of a mapping from being conformal. Therefore, by minimizing the integral in Eq.~\eqref{eqt:optimize_r}, we obtain an optimal disk radius $r^*$ such that the associated OMT map $h_{r^*} \circ r^*g$ is as conformal as possible. 

\subsubsection{Inverse stereographic projection}
Finally, we apply the inverse stereographic projection $\varphi^{-1}$ in Eq.~\eqref{eqt:stereographic_inverse} to map the planar OMT mapping result onto a spherical cap. The overall adaptive spherical cap parameterization is given by
\begin{equation} \label{eqt:overall_f}
    f = \varphi^{-1} \circ h_{r^*} \circ r^* g,
\end{equation}
Note that if $0<r^*<1$, $f$ maps the object surface $\mathcal{S}_o$ to a spherical cap which is smaller than a hemisphere. If $r^*>1$, $f$ maps $\mathcal{S}_o$ to a spherical cap larger than a hemisphere. The lower bound of the $Z$-value of the adaptive spherical cap is given by
\begin{equation}
    Z^* = \frac{1 - (r^*)^2}{1 + (r^*)^2}.
\end{equation}
We remark that the effect of $\varphi^{-1}$ on the area distortion has already been taken into account in the previous OMT mapping step, and hence $f$ is an area-preserving map. Also, since $\varphi^{-1}$ is conformal, this projection step does not affect the conformality of the previously optimized map $h_{r^*} \circ r^*g$. 

\subsection{Adaptive Area-Preserving Parameterization of Genus-0 Closed Surfaces}
Let $\mathcal{S}_c$ be a genus-0 closed surface. We compute an area-preserving map of $\mathcal{S}_c$ onto an optimal spherical cap domain $\mathbb{S}^2_{Z \geq Z^*}$ with the bottom part filled.

\subsubsection{Initial flattening map}
Similar to the open surface case, our strategy is to reduce the mapping problem to a problem on the plane. As the input surface is closed, it is necessary to puncture certain part of it in order to flatten it onto the plane. In the discrete case, we treat $\mathcal{S}_c$ as a triangular mesh and hence it is natural to consider puncturing a minimal set of triangular faces. Here, we follow the approach in~\cite{giri2021open} to puncture a quadrilateral region at the bottom part of the surface. More specifically, we first rotate the surface mesh to align it with the $Z$-axis and then find a pair of triangles $T_1, T_2$ sharing a common edge at the bottom part of it such that the quadrilateral formed by the four vertices of the triangles is as regular as possible. Starting from the pair of triangles closest to the center of the bottom part of the surface mesh, we first compute the length of the diagonals and the edges of the quadrilateral. Then, we consider the ratio of the maximum side length to the minimum side length, and the ratio of the length of the longer diagonal to that of the shorter diagonal. If both ratios are sufficiently close to 1, then the quadrilateral is considered to be regular. If the ratios are much greater than 1, we search for the next pair of triangles and repeat the above procedure. By puncturing $T_1$ and $T_2$, the surface becomes a topological disk and hence can be flattened on the plane.

As described in~\cite{giri2021open}, the disk conformal map~\cite{choi2015fast} may induce a large area distortion for this punctured surface as the punctured quadrilateral region is very small relative to the entire surface. Therefore, we follow the approach in~\cite{giri2021open} and combine the stretch energy minimization (SEM) method~\cite{yueh2019novel} and the quasi-conformal composition~\cite{choi2015flash} for obtaining the initial flattening map. The procedure is outlined below.

We first apply the SEM method to map $\mathcal{S}_c \setminus \{T_1, T_2\}$ onto the unit disk $\mathbb{D}$, with the four vertices of the punctured quadrilateral region mapped onto the unit circle. Denote the SEM map as $\tilde{g}$ and the associated Beltrami coefficient as $\mu_{\tilde{g}}$. To reduce the conformal distortion of $\tilde{g}$, we apply the quasi-conformal composition~\cite{choi2015flash} and reconstruct a quasi-conformal map $g: \mathcal{S}_c \setminus \{T_1, T_2\} \to \mathbb{C}$ with the Beltrami coefficient $\lambda \mu_{\tilde{g}}$, where $\lambda \in [0,1]$ is a scaling factor for controlling the conformal distortion. In practice, we set $\lambda = 0.2$ to achieve an initial map $g$ with moderate distortion in both area and angle.

\subsubsection{Optimization on the plane}
With the initial flattening map $g$ computed, we can now simply follow the procedure in the open surface case to optimize both the size of the disk and the mapping by solving the minimization problem in Eq.~\eqref{eqt:optimize_r}. Denote the optimal radius by $r^*$ and the OMT map as $h_{r^*}$.

\subsubsection{Inverse stereographic projection}
Finally, we apply the inverse stereographic projection $\varphi^{-1}$ to map the planar mapping result onto a spherical cap, thereby obtaining the overall adaptive area-preserving parameterization $f = \varphi^{-1} \circ h_{r^*} \circ r^* g$ with the punctured quadrilateral region filled. We remark that only the four vertices of the quadrilateral are mapped to the bottom circle $X^2 + Y^2 = 1- (Z^*)^2$. All other vertices of $\mathcal{S}_c$ are mapped to $\mathbb{S}^2_{Z > Z^*}$. 


\begin{figure}[t]
\centering
\begin{subfigure}[b]{0.15\textwidth}
\centering
\includegraphics[scale=0.2]{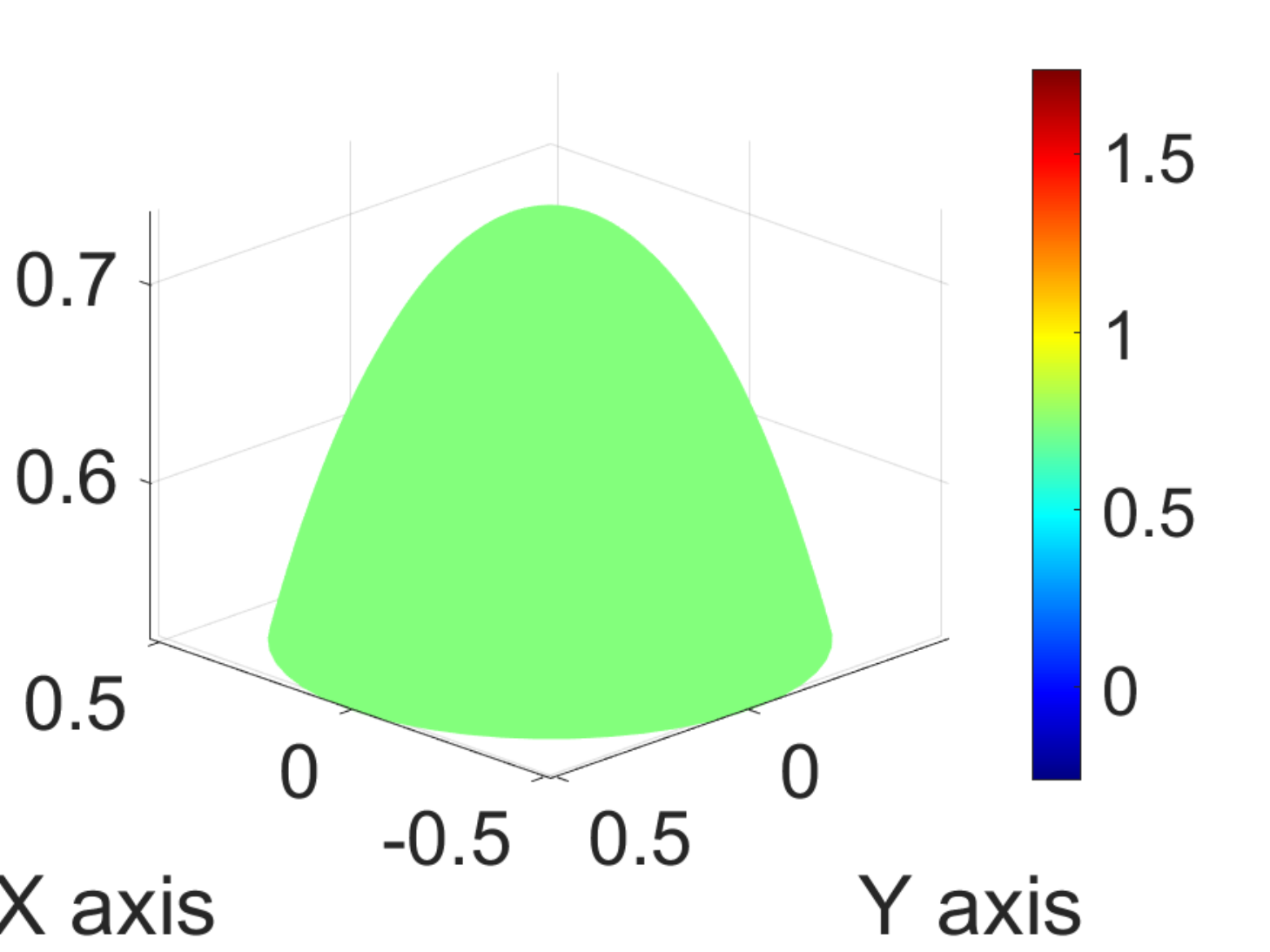}
\caption{$A_0^0$}
\label{AAH00}
\end{subfigure}
~~\begin{subfigure}[b]{0.15\textwidth}
\centering
\includegraphics[scale=0.20]{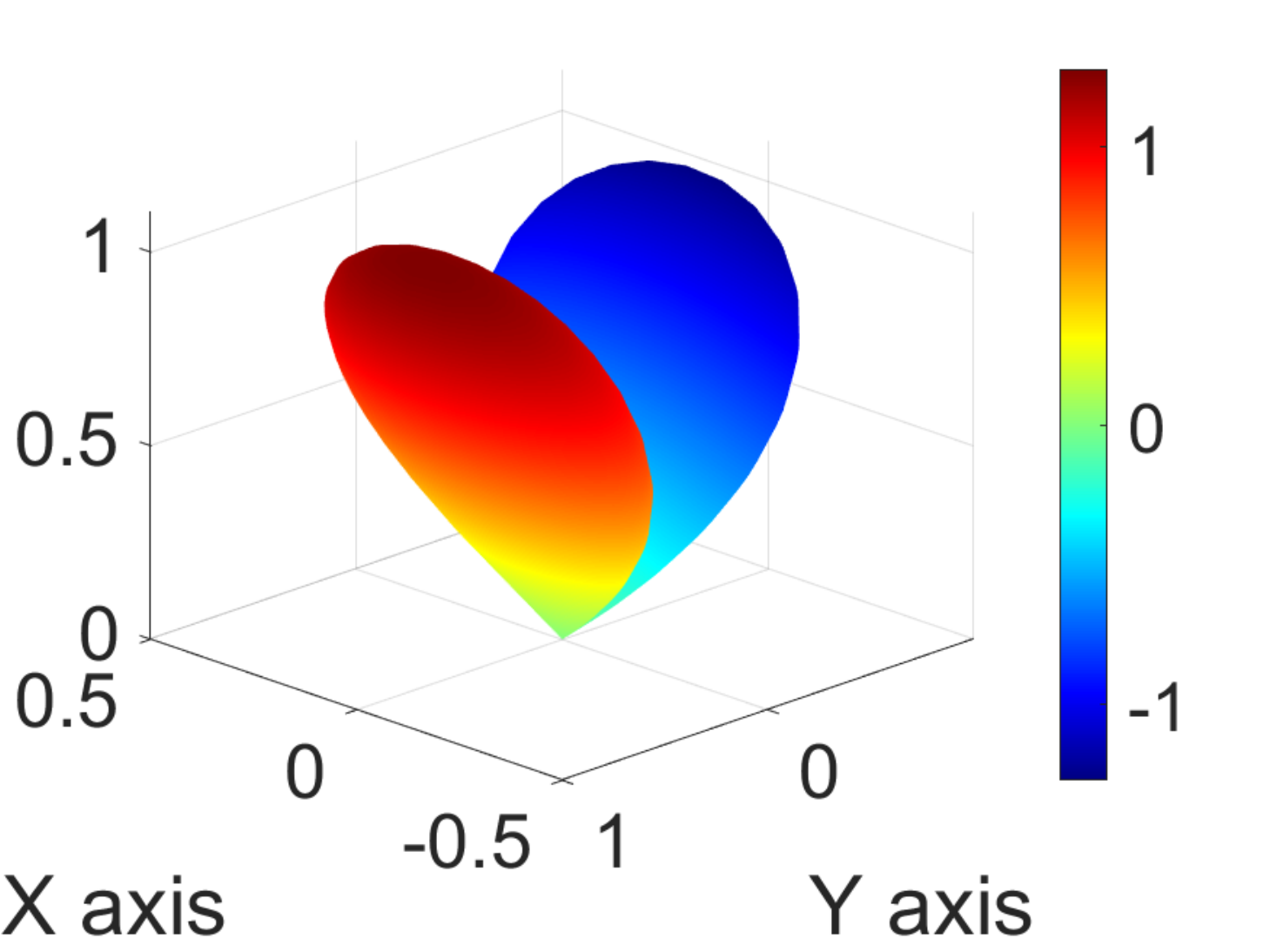}
\caption{$A_{1}^{-1}$}
\label{AAH1-1}
\end{subfigure}
~ \begin{subfigure}[b]{0.15\textwidth}
\centering
\includegraphics[scale=0.20]{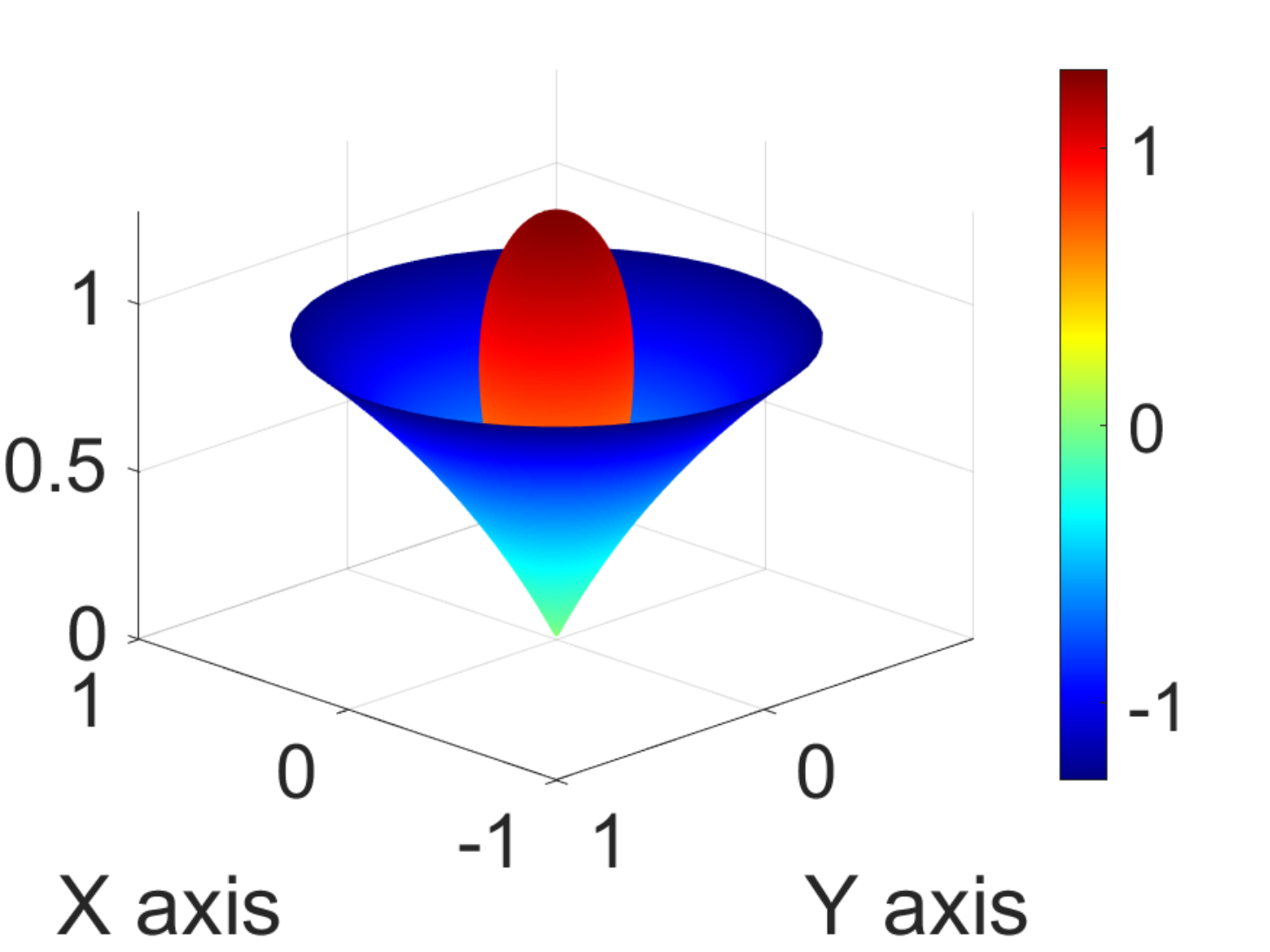}
\caption{$A_{1}^{0}$}
\label{AAH10}
\end{subfigure}
~ \begin{subfigure}[b]{0.15\textwidth}
\centering
\includegraphics[scale=0.20]{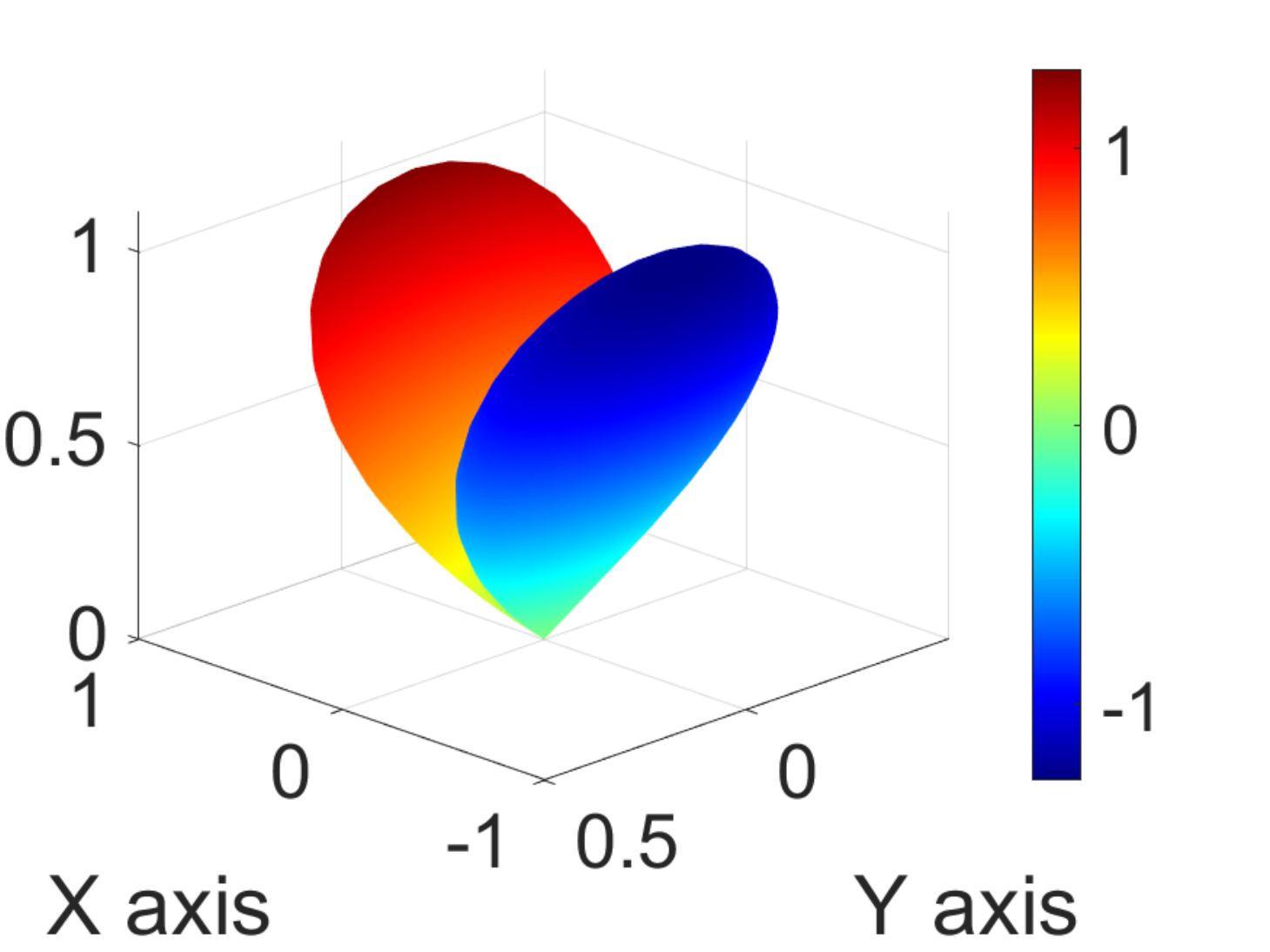}
\caption{$A_{1}^{1}$}
\label{AAH11}
\end{subfigure}
~\begin{subfigure}[b]{0.15\textwidth}
\centering
\includegraphics[scale=0.20]{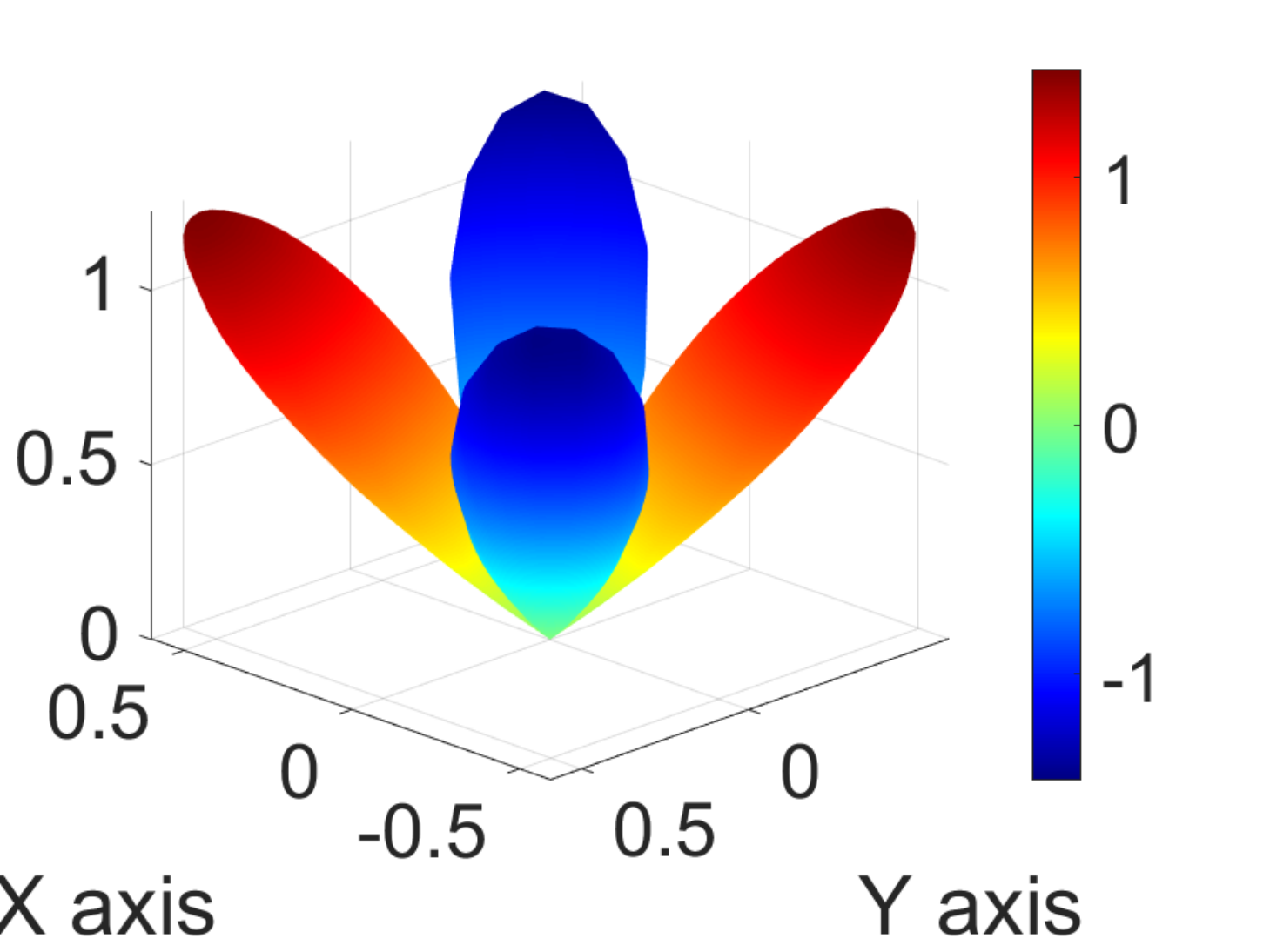}
\caption{$A_{2}^{-2}$}
\label{AAH2-2}
\end{subfigure}
~ \begin{subfigure}[b]{0.15\textwidth}
\centering
\includegraphics[scale=0.20]{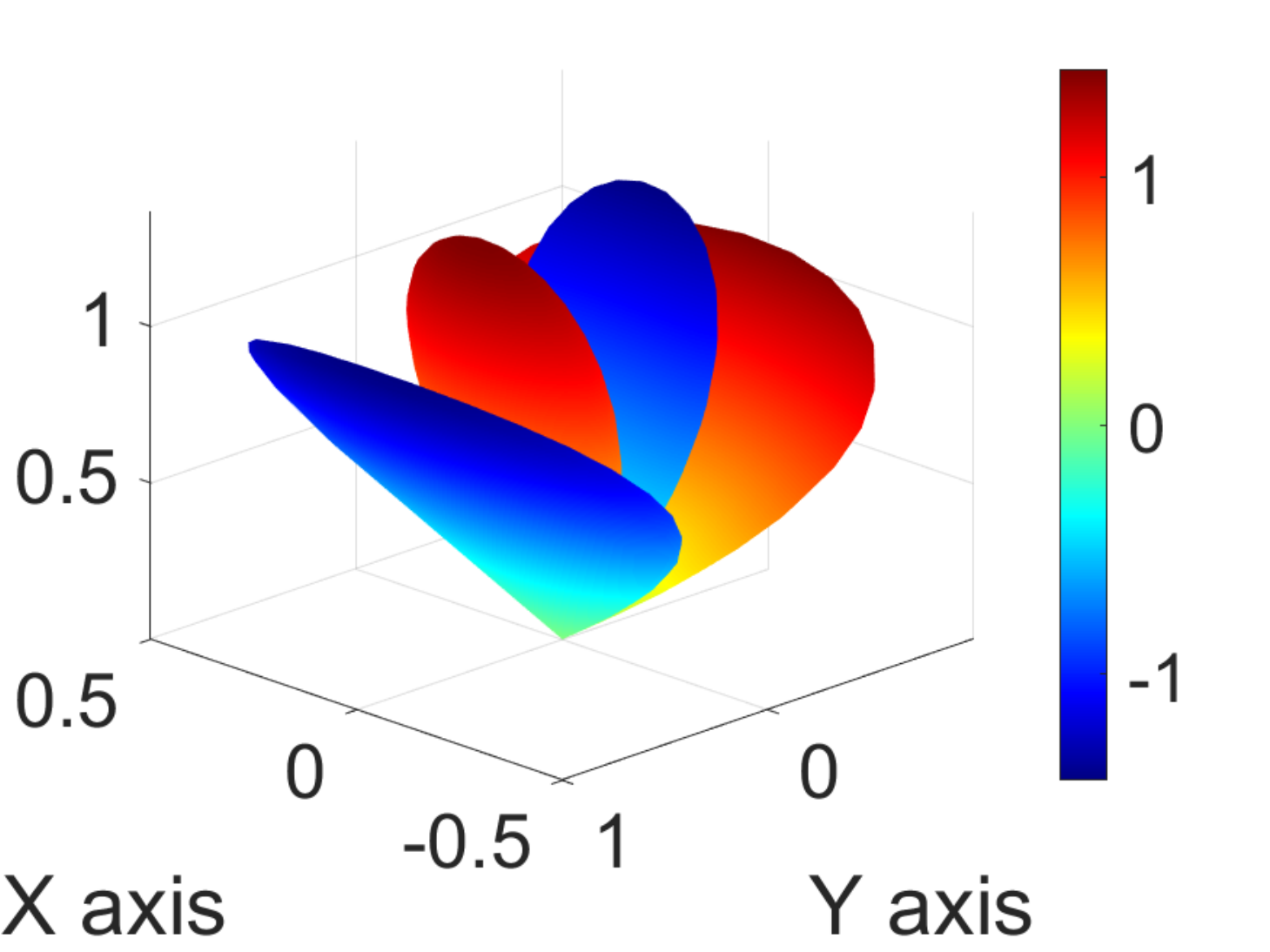}
\caption{$A_{2}^{-1}$}
\label{AAH2-1}
\end{subfigure}
~ \begin{subfigure}[b]{0.15\textwidth}
\centering
\includegraphics[scale=0.20]{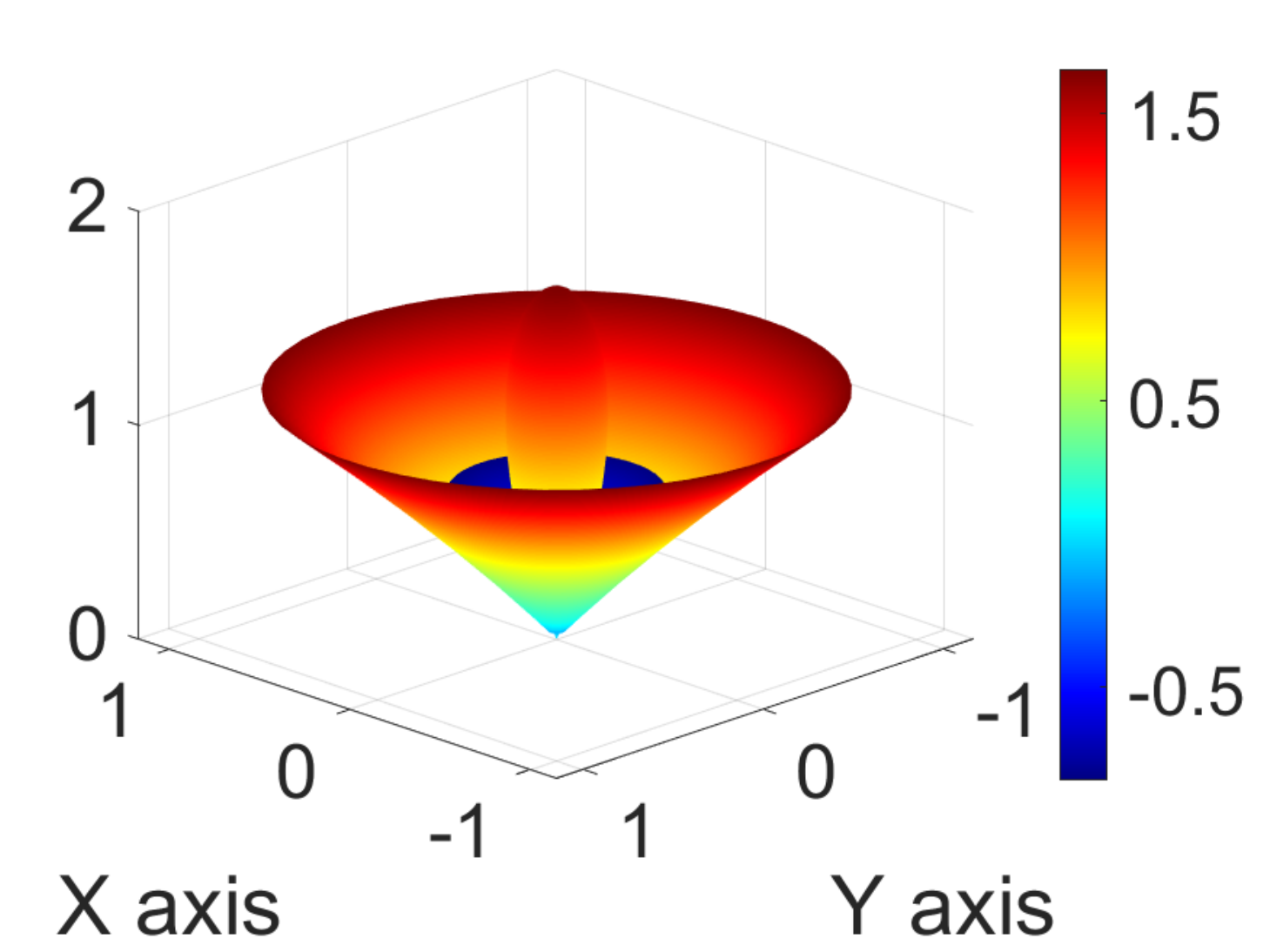}
\caption{$A_{2}^{0}$}
\label{AAH20}
\end{subfigure}
~\begin{subfigure}[b]{0.15\textwidth}
\centering
\includegraphics[scale=0.20]{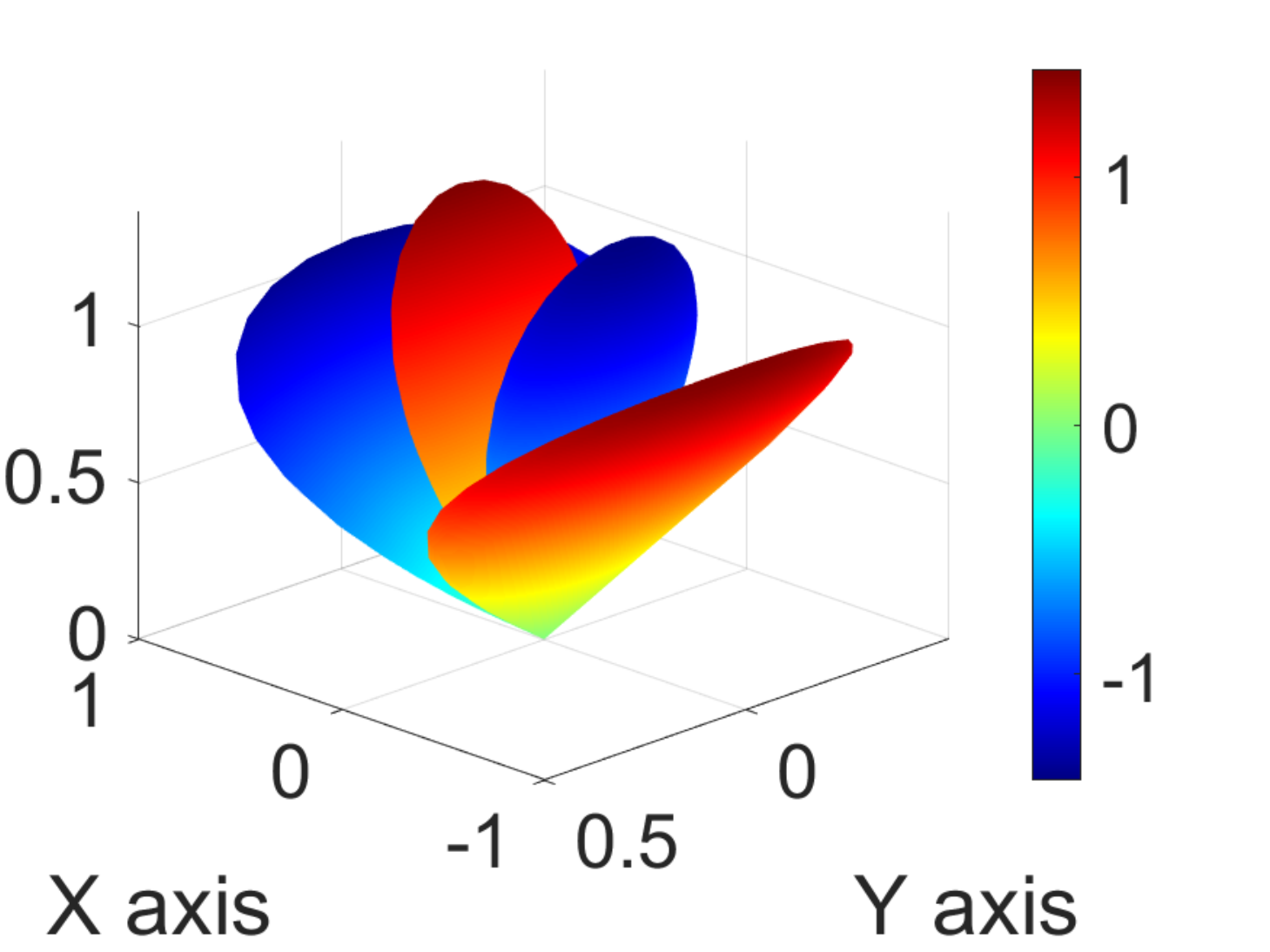}
\caption{$A_{2}^{1}$}
\label{AAH21}
\end{subfigure}
~ \begin{subfigure}[b]{0.15\textwidth}
\centering
\includegraphics[scale=0.20]{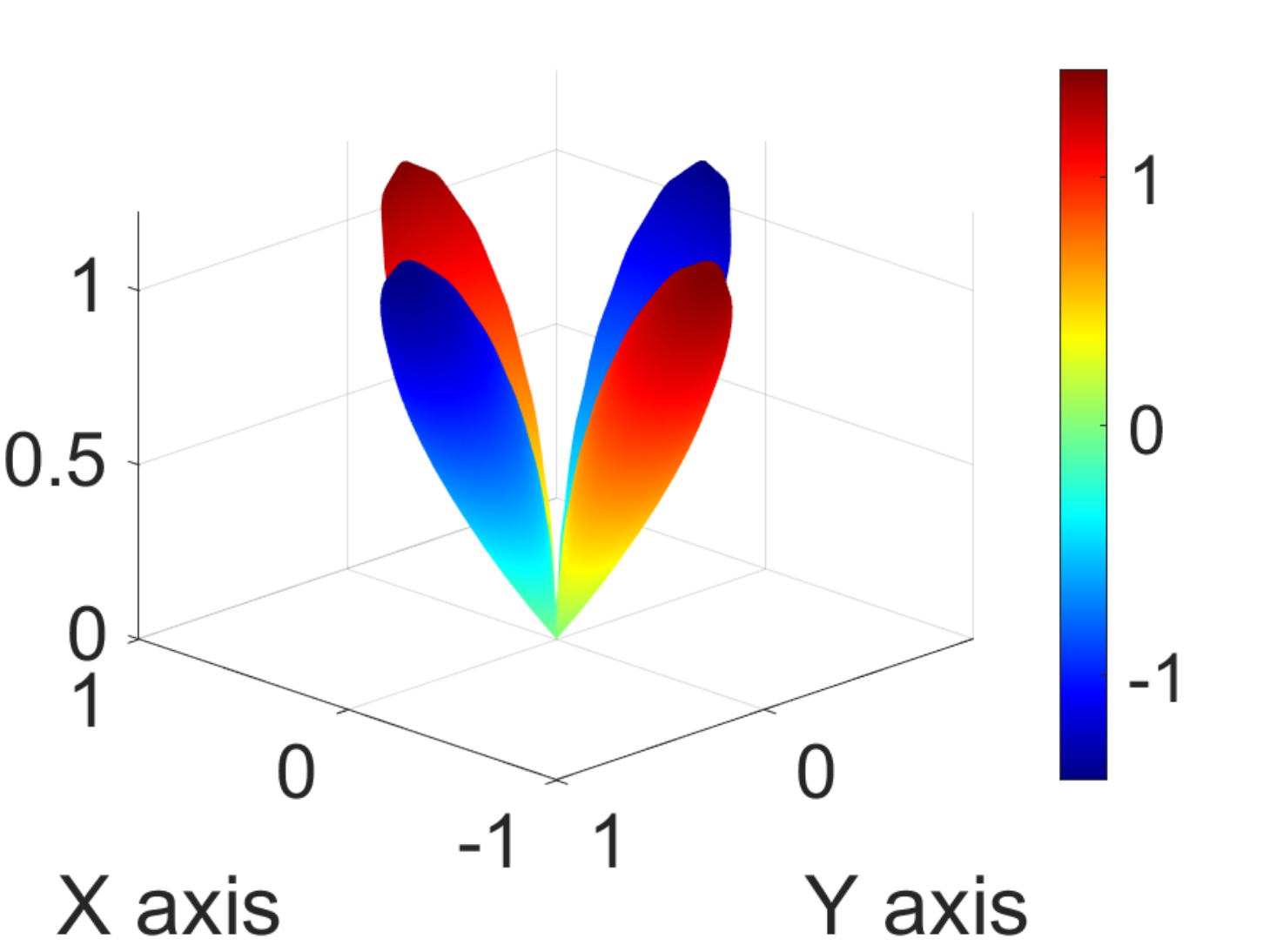}
\caption{$A_{2}^{2}$}
\label{AAH22}
\end{subfigure}
\caption{The AH basis functions up to the second order, with $Z^* > 0$. (a)~$n=0$. (b)-(d)~$n=1$. (e)-(i)~$n=2$ for less than a hemisphere ($\theta=\pi/4$). The distance between each surface point and origin indicates the magnitude of $A_{n}^{m}$, and the color represents the actual value of $A_{n}^{m}$.}
\label{fig:hshfig1}
\end{figure}

\begin{figure}[t]
\centering
\begin{subfigure}[b]{0.15\textwidth}
\centering
\includegraphics[scale=0.2]{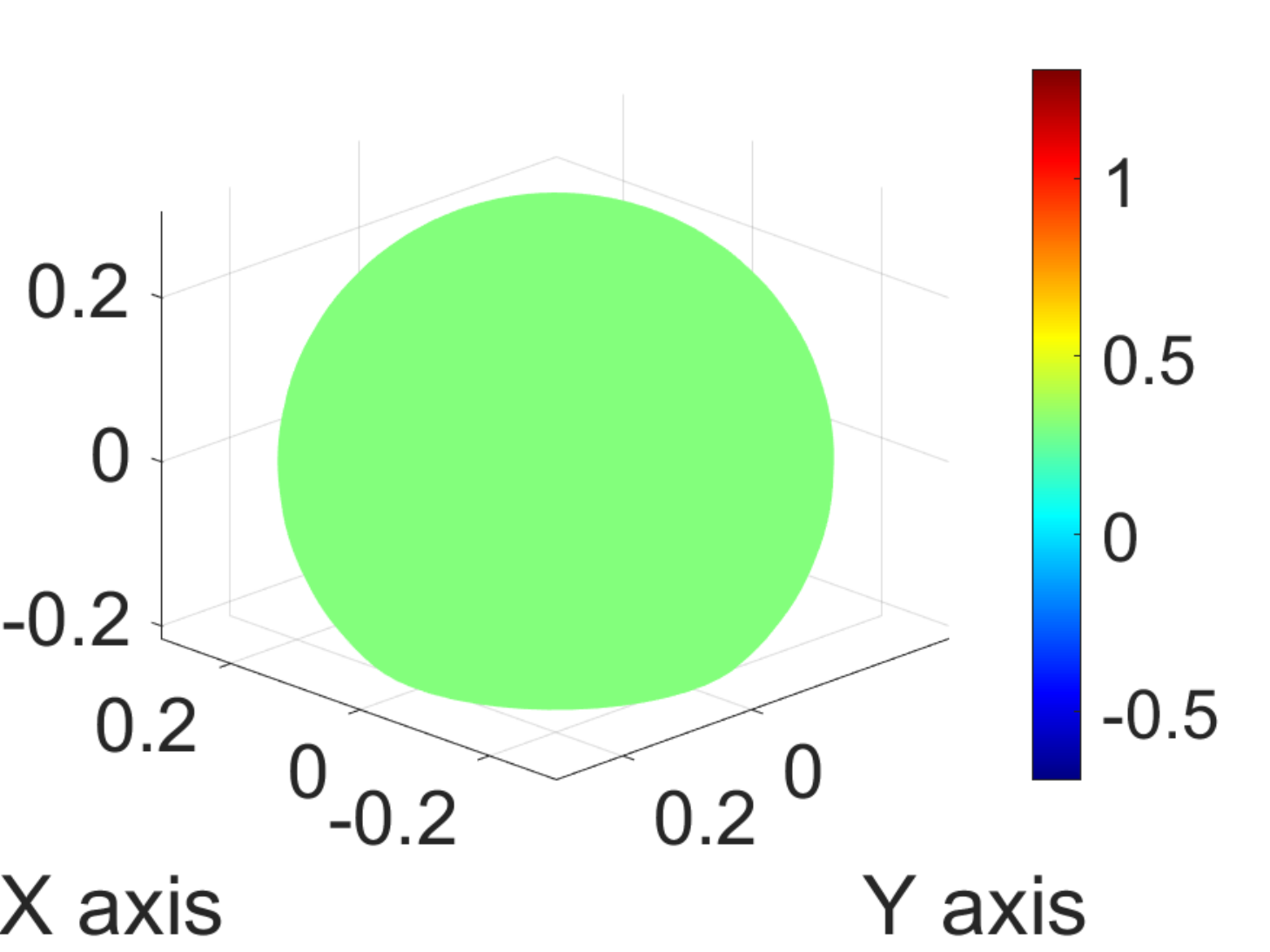}
\caption{$A_0^0$}
\label{AH00}
\end{subfigure}
~~\begin{subfigure}[b]{0.15\textwidth}
\centering
\includegraphics[scale=0.20]{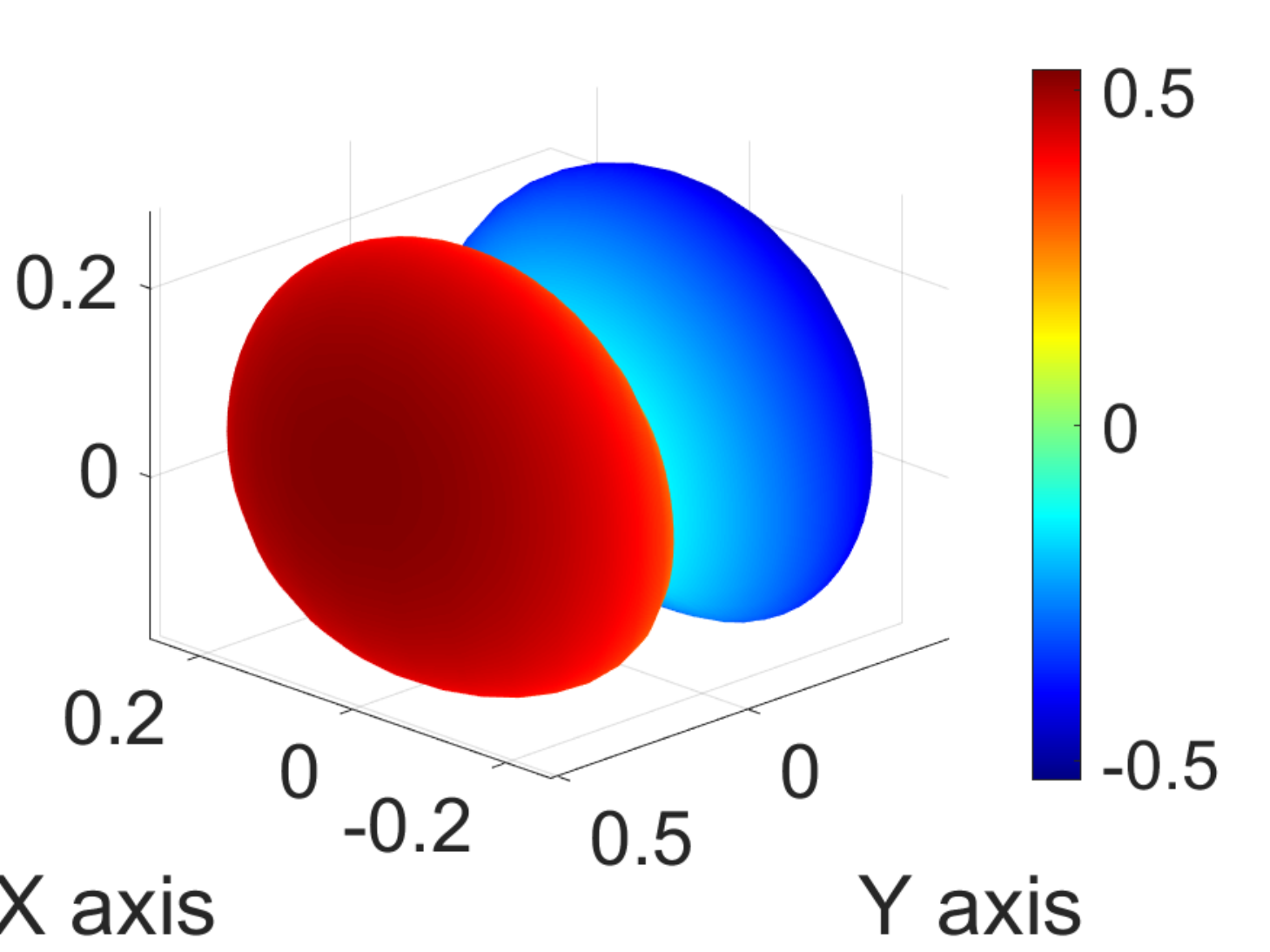}
\caption{$A_{1}^{-1}$}
\label{AH1-1}
\end{subfigure}
~ \begin{subfigure}[b]{0.15\textwidth}
\centering
\includegraphics[scale=0.20]{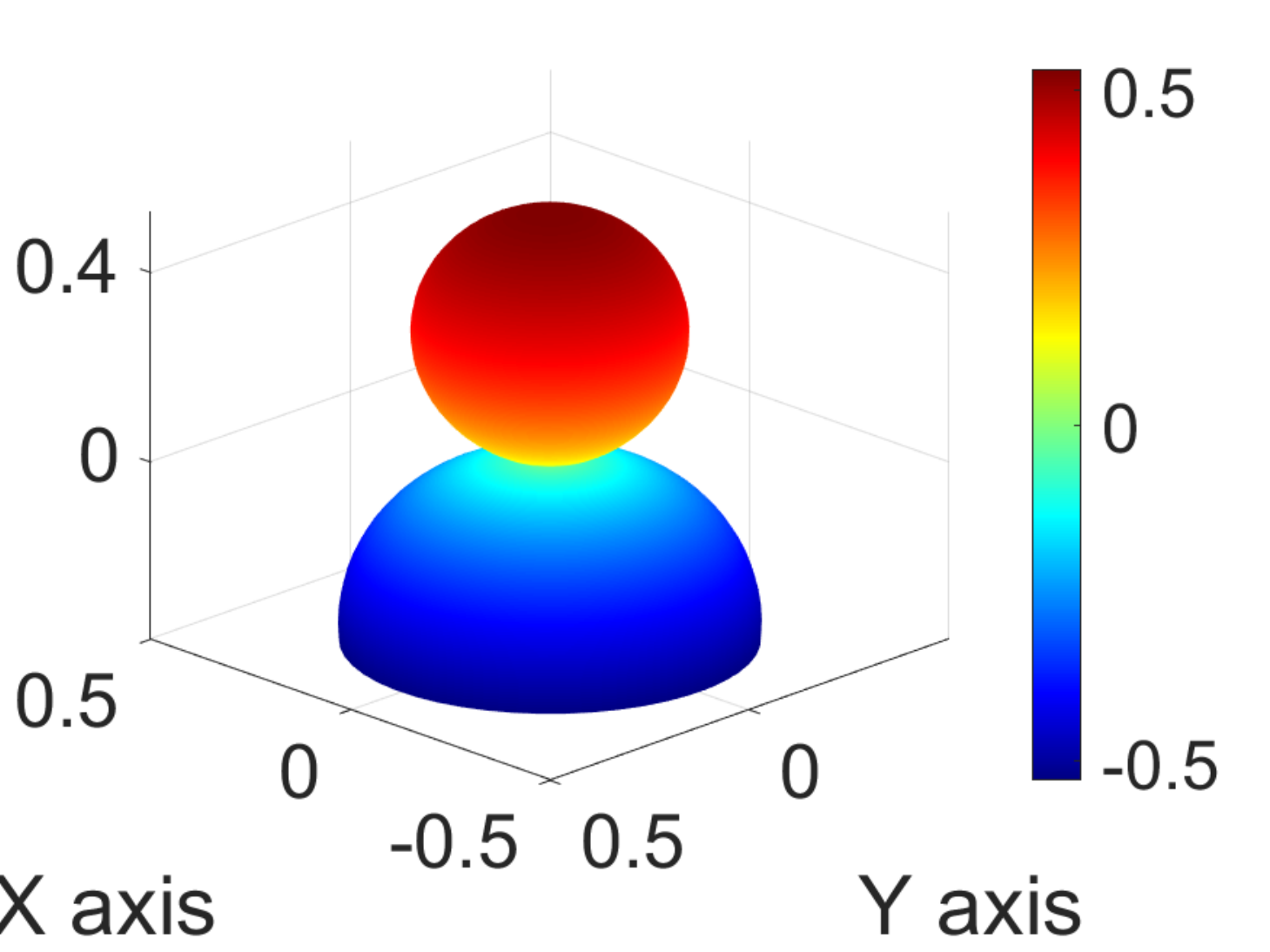}
\caption{$A_{1}^{0}$}
\label{AH10}
\end{subfigure}
~ \begin{subfigure}[b]{0.15\textwidth}
\centering
\includegraphics[scale=0.20]{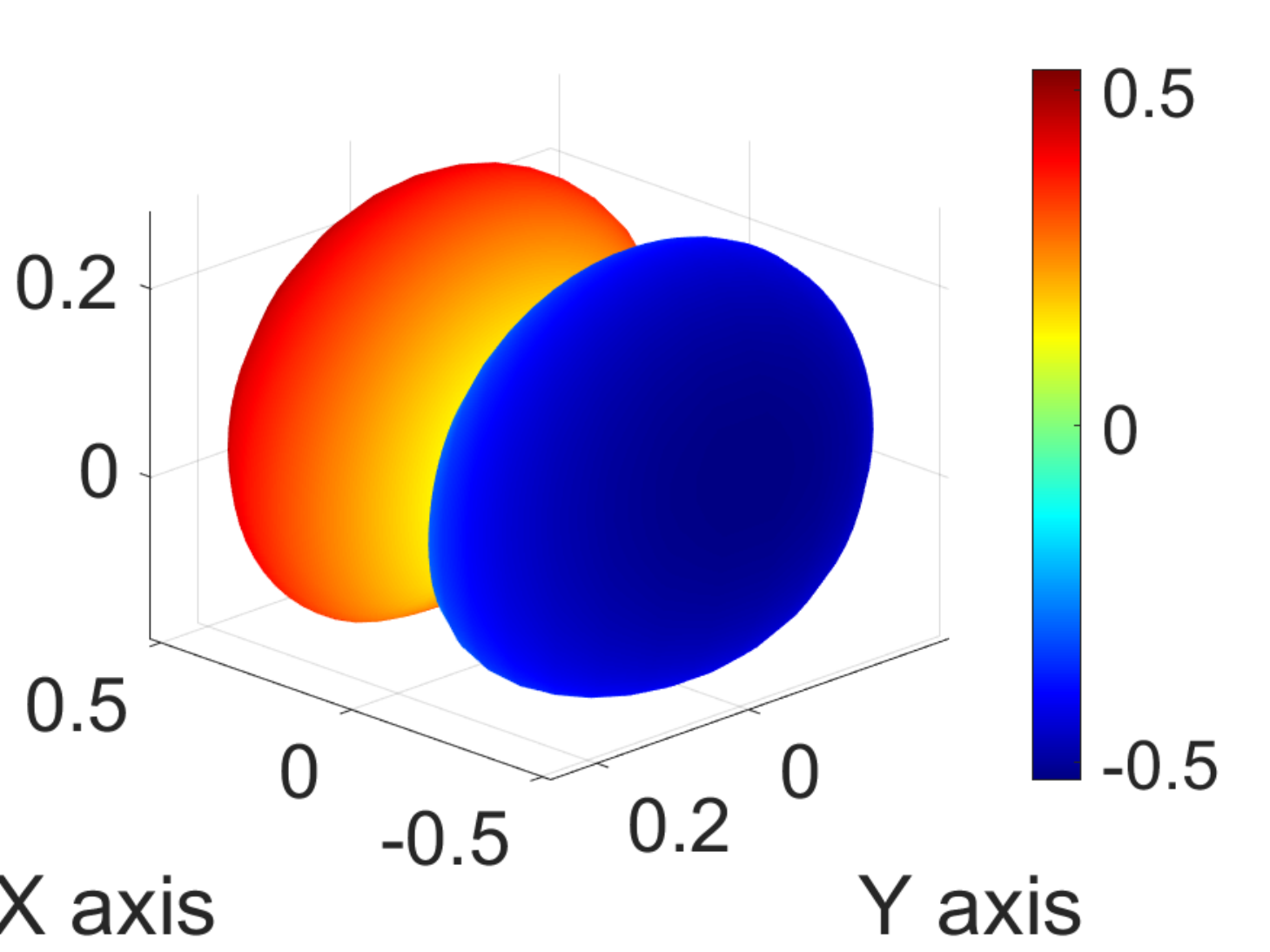}
\caption{$A_{1}^{1}$}
\label{AH11}
\end{subfigure}
~\begin{subfigure}[b]{0.15\textwidth}
\centering
\includegraphics[scale=0.20]{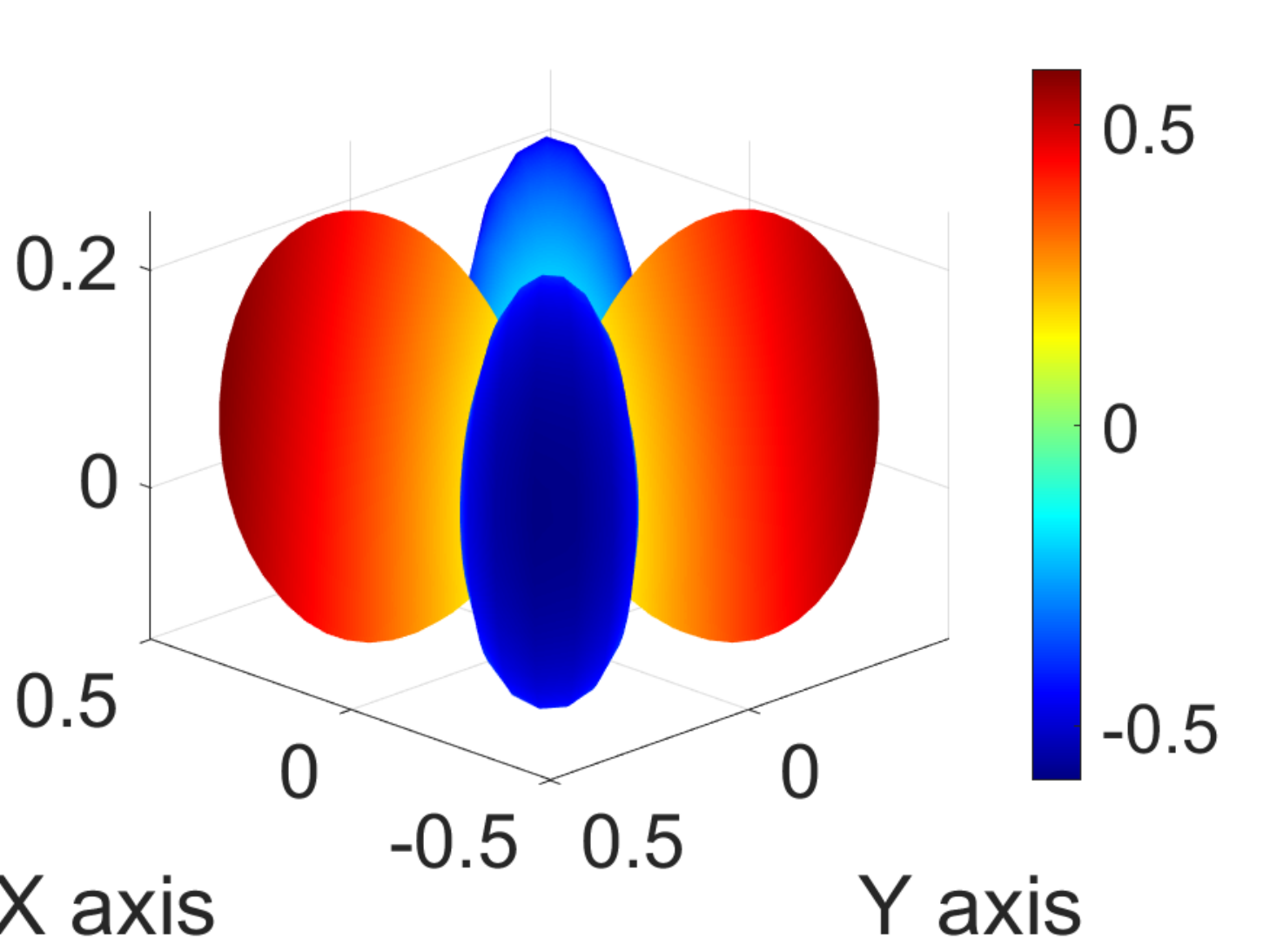}
\caption{$A_{2}^{-2}$}
\label{AH2-2}
\end{subfigure}
~ \begin{subfigure}[b]{0.15\textwidth}
\centering
\includegraphics[scale=0.20]{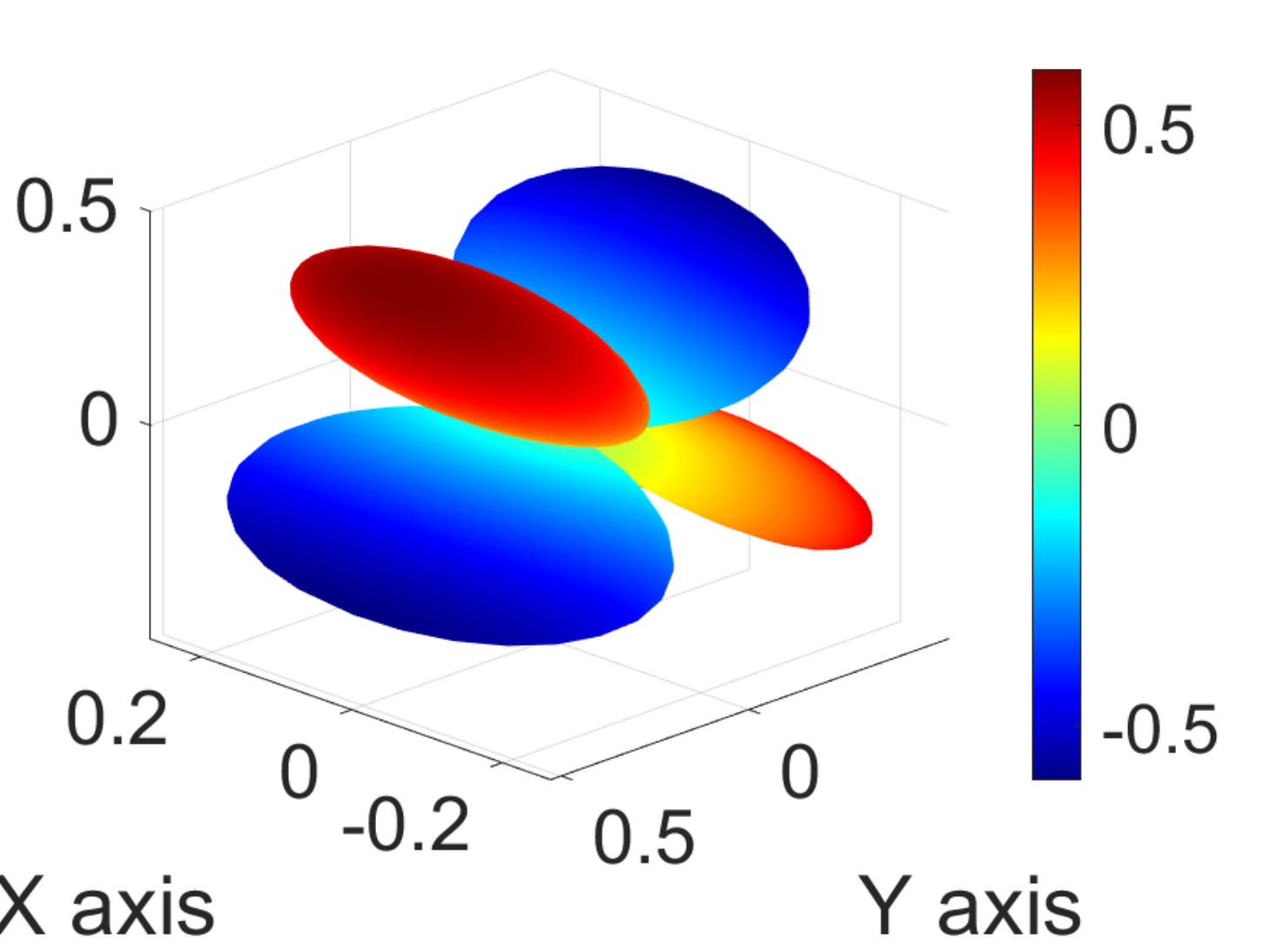}
\caption{$A_{2}^{-1}$}
\label{AH2-1}
\end{subfigure}
~ \begin{subfigure}[b]{0.15\textwidth}
\centering
\includegraphics[scale=0.20]{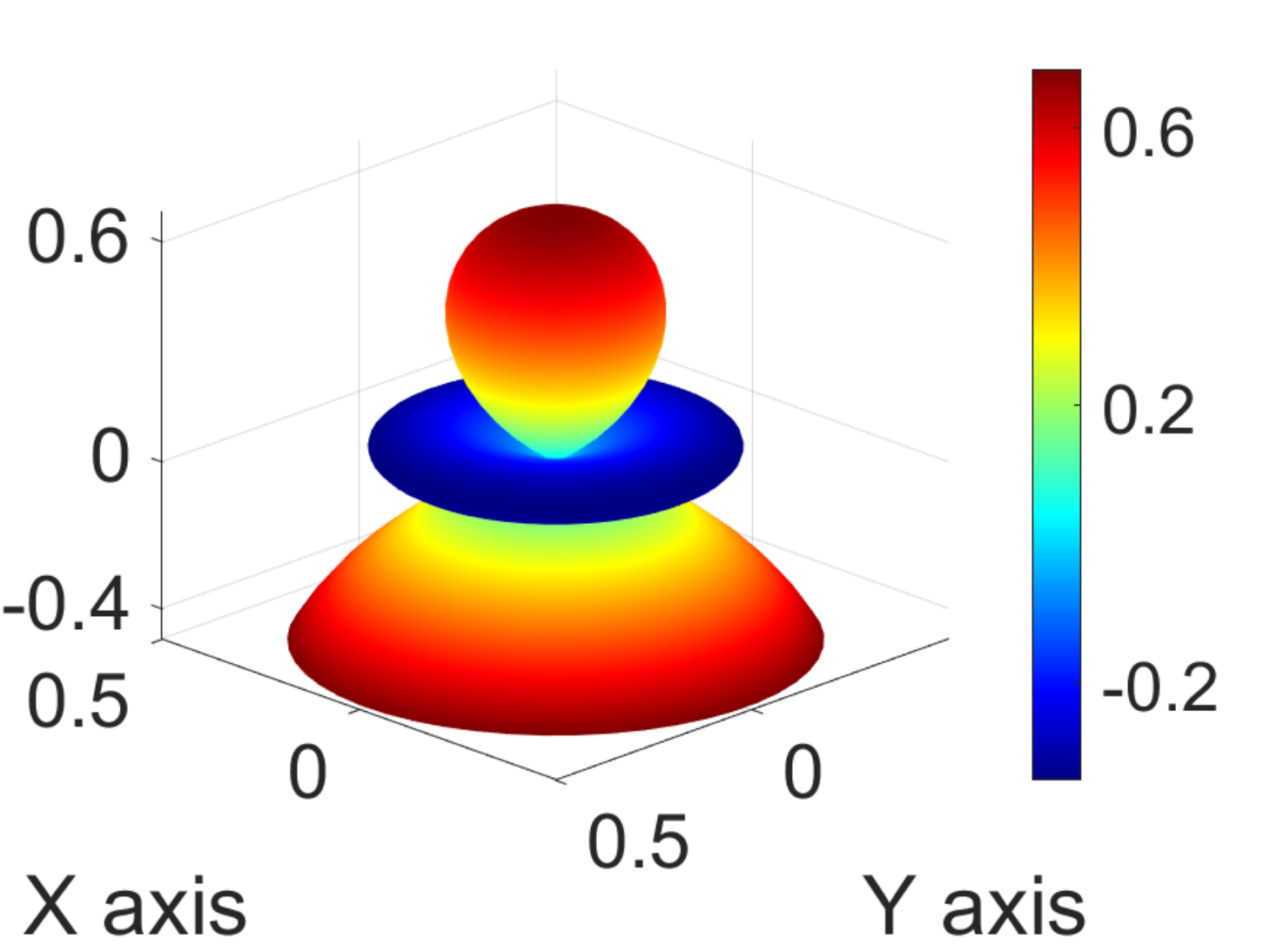}
\caption{$A_{2}^{0}$}
\label{AH20}
\end{subfigure}
~\begin{subfigure}[b]{0.15\textwidth}
\centering
\includegraphics[scale=0.20]{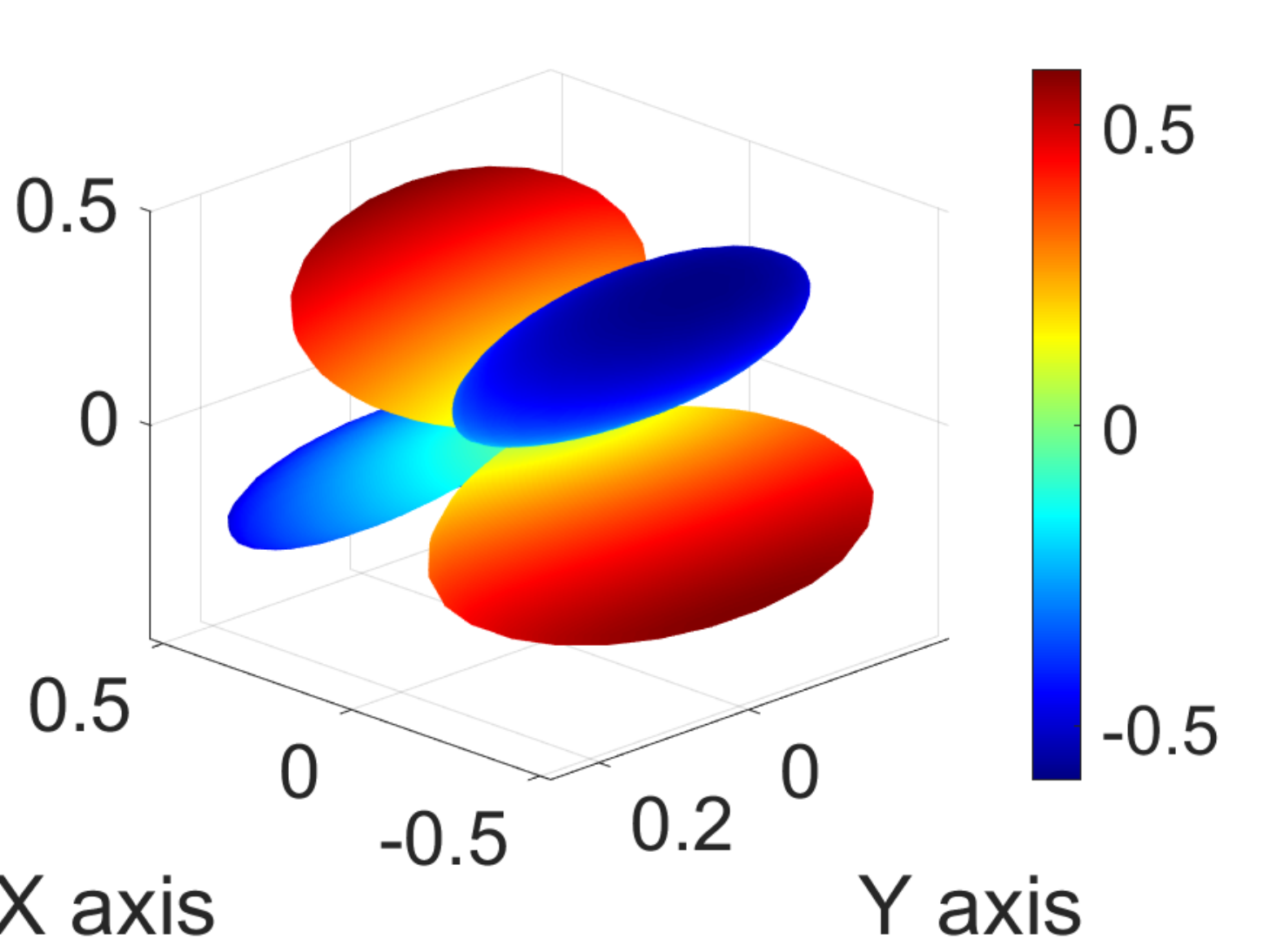}
\caption{$A_{2}^{1}$}
\label{AH21}
\end{subfigure}
~ \begin{subfigure}[b]{0.15\textwidth}
\centering
\includegraphics[scale=0.20]{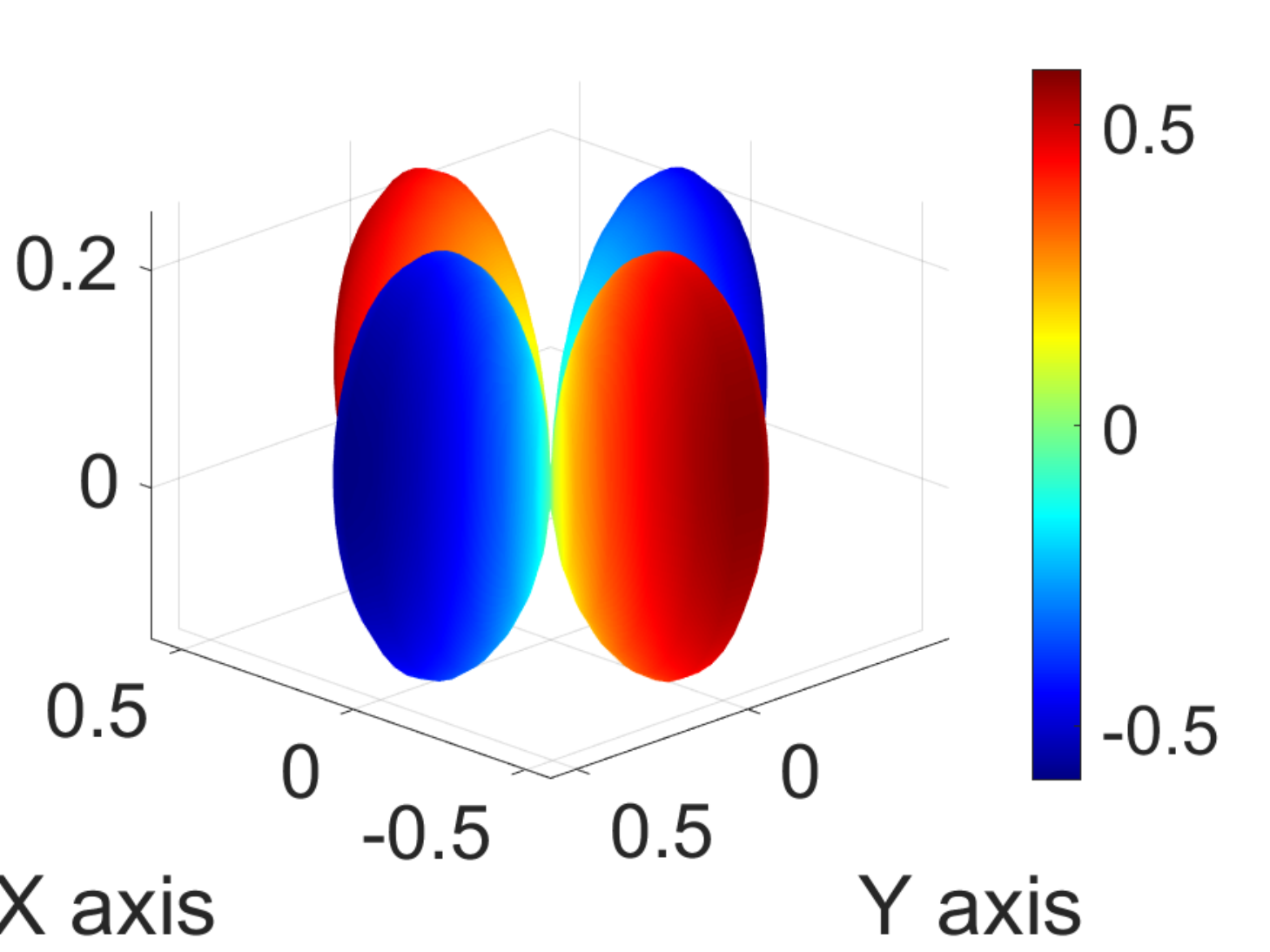}
\caption{$A_{2}^{2}$}
\label{AH22}
\end{subfigure}
\caption{The AH basis functions up to the second order, with $Z^* < 0$. (a)~$n=0$. (b)-(d)~$n=1$. (e)-(i)~$n=2$ for more than a hemisphere ($\theta=3\pi/4$). The distance between each surface point and origin indicates the magnitude of $A_{n}^{m}$, and the color represents the actual value of $A_{n}^{m}$.}
\label{fig:hshfig2}
\end{figure}
\subsection{Adaptive Harmonics (AH)}
In~\cite{huang2006hemispherical}, Huang \emph{et al.} extended the concept of spherical harmonics (SH) and developed a set of hemispherical harmonics (HSH) basis functions over the unit hemisphere, which have been found useful for brain source localization~\cite{giri2018eeg,giri2018ohbm} and surface description~\cite{giri2021open}. More recently, Giri \emph{et al.}~\cite{giri2019head,giri2020brain} developed an extension of HSH called the head harmonics (H$^2$) specifically as per the human head dimension. Here, we develop AH, a more general set of SH-like basis functions defined over the adaptive spherical cap region $\mathbb{S}^2_{Z \geq Z^*}$, by further extending the formulation of SH and HSH. The resulting AH will be utilized for efficient anatomical shape description and reconstruction.

For any order $n \geq 0$ and degree $m \in [-n, n]$, the associated Legendre polynomial (ALP) $P_{n}^{m}$ is defined as
\begin{equation} 
P_{n}^{m}(x)=\frac{(-1)^m}{2^nn!}(1-x^2)^{{m/2}}\frac{\mathrm{d}^{n+m}}{\mathrm{d}x^{n+m}}(x^2-1)^n.
\end{equation}
It is easy to see that for any fixed degree $m$, the ALPs of different orders are all orthogonal over $x\in[-1,1]$ (see~\cite{sharmonics_mathbook} for details). As described in~\cite{szeg1939orthogonal}, for any $q_1, q_2$ with $q_1 \neq 0$, the shifted ALPs $\widetilde P_{n}^{m}(x) = P_{n}^{m} (q_{1}x+q_2)$ are then orthogonal over the interval $\left[\frac{-1-q_2}{q_1},\frac{1-q_2}{q_1}\right]$. Now, note that the adaptive spherical cap region $\mathbb{S}^2_{Z \geq Z^*}$ can be expressed as $\{(\theta, \phi): \theta \in [0, \theta^*], \phi \in [-\pi, \pi]\}$, where $\theta$ is the elevation angle and $\phi$ is the azimuth angle. The upper limit of the elevation angle can be written as $\theta^* = \cos^{-1} Z^*$. For the shifted ALPs $\widetilde P_{n}^{m}$ to be orthogonal over the interval $[Z^* , 1]$, we have 
\begin{equation}
\left\{\begin{array}{ll}
    q_1 Z^* + q_2  &= -1,  \\
    q_1 + q_2 &= 1, 
\end{array} \right.
\end{equation}
which gives
\begin{equation}
\begin{pmatrix}
q_1 \\ q_2
\end{pmatrix} = \begin{pmatrix}
Z^* & 1 \\ 1 & 1
\end{pmatrix}^{-1} \begin{pmatrix}
-1 \\ 1
\end{pmatrix} = \frac{1}{Z^*-1} \begin{pmatrix}
2 \\ Z^*+1
\end{pmatrix}.
\end{equation}
The orthogonality relation of the shifted ALPs for the above $q_1, q_2$ is then given by
\begin{align}
\int_{0}^{1} \widetilde P_{n}^{m}(x)\widetilde P_{n'}^m(x)dx = \frac{2\left( n+m\right) !}{q_1\left( 2n+1\right) \left( n-m\right) !} \delta_{nn'},
\label{eq:22}
\end{align}
where $\delta_{nn'}$ is the Kronecker delta function. Now, the shifted ALPs can be utilized for constructing the AH basis functions over the adaptive spherical cap $\mathbb{S}^2_{Z \geq Z^*}$. The real-valued AH functions $A_n^m: [0,\cos^{-1} Z^*] \times [-\pi,\pi] \to \mathbb{R}$ for $n \geq 0$ and $m \in [-n, n]$ are defined as follows:
\begin{equation}
    \resizebox{\hsize}{!}{$
  A_n^m(\theta,\phi) = \left\{
   \begin{array}{lll}
    (-1)^{|m|}\sqrt{2} \widetilde K_{n}^{m} \sin(|m|\phi)\widetilde P_{n}^{|m|} (\cos\theta)
    & : m < 0, \\
    (-1)^{|m|}\sqrt{2}\widetilde K_{n}^{m} \cos(m\phi)\widetilde P_{n}^{m}(\cos\theta) & : m > 0,\\
    \widetilde K_{n}^{0}\widetilde P_{n}^{0}(\cos\theta) & : m=0,  
   \end{array}
   \right. 
   $}
\end{equation}
where $\widetilde K_{n}^{m}$ is a normalization constant with
\begin{equation}
\widetilde K_{n}^{m} = \sqrt{\frac{q_1(2n+1)(n-|m|)!}{4\pi (n+|m|)!}}.
\label{eq:24}
\end{equation}
Fig.~\ref{fig:hshfig1} and Fig.~\ref{fig:hshfig2} show the AH basis functions up to the second order with $Z^* > 0$ and $Z^* < 0$ respectively. 

The AH basis functions provide an effective way for the representation of different shapes. More specifically, given any open or closed anatomical surface with $k$ vertices, we can first compute the adaptive area-preserving parameterization $f$ onto an optimal spherical cap. Each point $v = (X,Y,Z)$ on the object surface is associated with a unique pair $(\theta,\phi) = \left(\cos^{-1}{\frac{Z_{f}}{\sqrt{X_{f}^2+Y_{f}^2+Z_{f}^2}}}, \tan^{-1} {\frac{Y_{f}}{X_{f}}}\right)$, where $(X_f, Y_f, Z_f) = f(X,Y,Z)$. Analogous to the SH~\cite{brechbuhler1995parametrization} and HSH~\cite{huang2006hemispherical} shape description, the object surface can be expressed as a weighted sum of the AH basis functions as
\begin{equation}
v(\theta,\phi) = \sum_{n=0}^{\infty} \sum_{m=-n}^{n}C_{n}^{m} A_{n}^{m}(\theta,\phi).
\end{equation}
In practice, for any prescribed maximum order $N$, we can approximate $v(\theta,\phi)$ using the AH basis functions up to order~$N$:
\begin{equation}\label{eqt:AH_reconstruction}
v(\theta,\phi) \approx \sum_{n=0}^{N} \sum_{m=-n}^{n}C_{n}^{m} A_{n}^{m}(\theta,\phi).
\end{equation}
The above can be further rewritten as a matrix equation $\mathbf{V} = \mathbf{A} \mathbf{C}$ where $\mathbf{V}$ is a $k\times 3$ matrix of the coordinates of all $k$ vertices, $\mathbf{A}$ is a $k \times (N+1)^2$ matrix of the AH basis functions, and $\mathbf{C} = (C_{0}^{0},C_{1}^{-1},C_{1}^{0}, \cdots, C_{N}^{N})^{T}$ is the AH coefficient matrix which can be estimated using the Moore-Penrose pseudo-inverse
\begin{equation} \label{eqt:inverse}
\mathbf{C} = (\mathbf{A}^T \mathbf{A})^{-1} \mathbf{A}^T \mathbf{V}.
\end{equation}

\begin{figure}[t]
    \centering
    \includegraphics[width=0.48\textwidth]{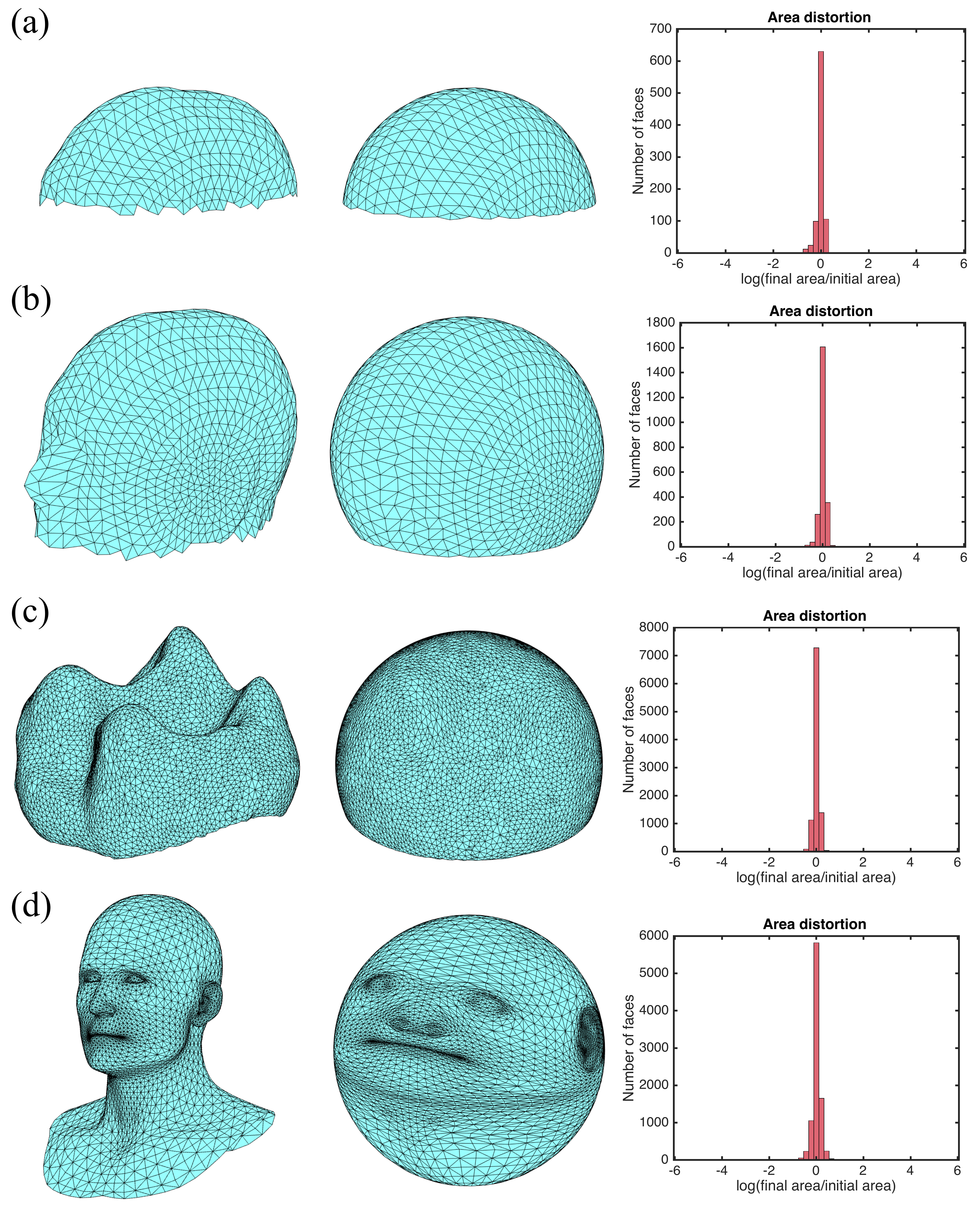}
    \caption{Adaptive area-preserving parameterization of simply-connected open surfaces obtained by our proposed algorithm. (a)~A small portion of a human scalp surface reconstructed from MRI images in the OASIS dataset~\cite{marcus2010open}. (b)~A larger portion of a human scalp surface reconstructed from MRI images in the OASIS dataset~\cite{marcus2010open}. (c)~A mammalian tooth surface from MorphoSource~\cite{winchester2014dental,gao2015hypoelliptic}. (d)~A human face from the CGTrader repository~\cite{CGTrader}. For each example, the input surface, the adaptive area-preserving parameterization and the area distortion histogram are shown.}
    \label{fig:mapping_result_open}
\end{figure}
\begin{table}[t!]
    \centering
    \begin{tabular}{c|c|c|C{22mm}}
    \multirow{ 2}{*}{Surface} & \multicolumn{3}{c}{mean$(|d_\text{area}|)$ / mean$(|d_\text{angle}|)$} \\ \cline{2-4}
         & Adaptive & Disk & Hemispherical\\ \hline
         Fig.~\ref{fig:mapping_result_open}(a) & 0.09 / 0.11 & 0.09 / 0.20 & 0.10 / 0.12\\
         Fig.~\ref{fig:mapping_result_open}(b) & 0.09 / 0.11 & 0.10 / 0.36 & 0.09 / 0.21\\
         Fig.~\ref{fig:mapping_result_open}(c) & 0.09 / 0.18 & 0.11 / 0.31 & 0.09 / 0.20\\
         Fig.~\ref{fig:mapping_result_open}(d) & 0.12 / 0.35 & 0.15 / 0.50 & 0.12 / 0.43
    \end{tabular}
    \caption{The performance of different methods for parameterizing simply-connected open surfaces.}
    \label{tab:performance_open}
\end{table}

\section{Experiments}\label{sect:experiment}
The proposed algorithms are implemented in MATLAB. The optimization problem~\eqref{eqt:optimize_r} is solved using the MATLAB built-in one-dimensional minimizer \texttt{fminbnd}. All experiments are performed on a PC with an Intel i7-6700K quad-core CPU and 16~GB RAM.

\subsection{Adaptive Area-Preserving Parameterization}
To demonstrate the flexibility of our proposed parameterization algorithms, we test them using various open and closed anatomical surfaces with different geometry. To assess the quality of the parameterizations, we evaluate the area distortion of the parameterization $f$ for any triangular face $T$ of the input surface $\mathcal{S}$ as follows:
\begin{equation}
d_{\text{area}}(T) = \log_e \frac{{\text{Area($f(T)$)}} / \left({\sum_{T' \in \mathcal{F}}\text{Area($f(T')$)}}\right)}{{\text{Area($T$)}} / \left({\sum_{T' \in \mathcal{F}}\text{Area($T'$)}}\right)},
\end{equation}
where $\mathcal{F}$ is the set of all triangular faces of $\mathcal{S}$. Note that the two summation terms are used for normalizing the total area of $\mathcal{S}$ and that of the resulting parameter domain so that the measure is nondimensionalized. For an ideal area-preserving parameterization, we should have $d_{\text{area}} \equiv 0$. We also consider the angle distortion of $f$, defined by the difference between any angle of any triangular face of $\mathcal{S}$ and the corresponding angle (in radian) in the resulting parameter domain:
\begin{equation}
d_{\text{angle}}([v_i, v_j, v_k]) = \angle [f(v_i), f(v_j), f(v_k)] - \angle  [v_i, v_j, v_k],
\end{equation}
where $[v_i, v_j, v_k]$ denotes the angle formed by the three vertices $v_i, v_j, v_k$ of $\mathcal{S}$. For an ideal conformal parameterization, we should have $d_{\text{angle}} \equiv 0$. 

\subsubsection{Parameterization of Open Anatomical Surfaces}
For simply-connected open anatomical surfaces, we first consider human scalp surfaces reconstructed from MRI images in the Open Access Series of Imaging Studies (OASIS) dataset~\cite{marcus2010open} (see~\cite{giri2021open} for more details of the reconstruction). As shown in Fig.~\ref{fig:mapping_result_open}(a)--(b), our method is capable of parameterizing surfaces with significantly different geometry. For the smaller portion of a human scalp in Fig.~\ref{fig:mapping_result_open}(a), the adaptive area-preserving parameterization gives a spherical cap domain less than a hemisphere. For the larger portion of a human scalp in Fig.~\ref{fig:mapping_result_open}(b), the parameterization gives a spherical cap domain greater than a hemisphere. In both cases, it can be observed from the distortion histograms that the parameterizations are highly area-preserving. In the next example, we consider a mammalian tooth from the biological data archive MorphoSource~\cite{winchester2014dental,gao2015hypoelliptic} (see Fig.~\ref{fig:mapping_result_open}(c)). Again, it can be observed that the adaptive parameterization is highly area-preserving. Finally, we consider a human face model freely available at the CGTrader repository~\cite{CGTrader} under the Royalty Free License (see Fig.~\ref{fig:mapping_result_open}(d)). While the face model consists of prominent features such as eyes and ears with relatively complicated geometry, our method successfully parameterizes the model onto an adaptive spherical cap domain with the area of the features well-preserved.

After demonstrating the effectiveness of our proposed adaptive area-preserving parameterization algorithm for handling different open anatomical surfaces, we compare it with the existing area-preserving parameterization methods including the disk OMT map~\cite{zhao2013area} and the hemispherical area-preserving parameterization~\cite{giri2021open} (i.e. with $Z^* = 0$). As shown in Table~\ref{tab:performance_open}, with the flexibility of the parameter domain, our adaptive area-preserving method can achieve area distortion comparable to or even smaller than the two other methods while effectively reducing the angle distortion. 

\begin{figure}[t!]
    \centering
    \includegraphics[width=0.48\textwidth]{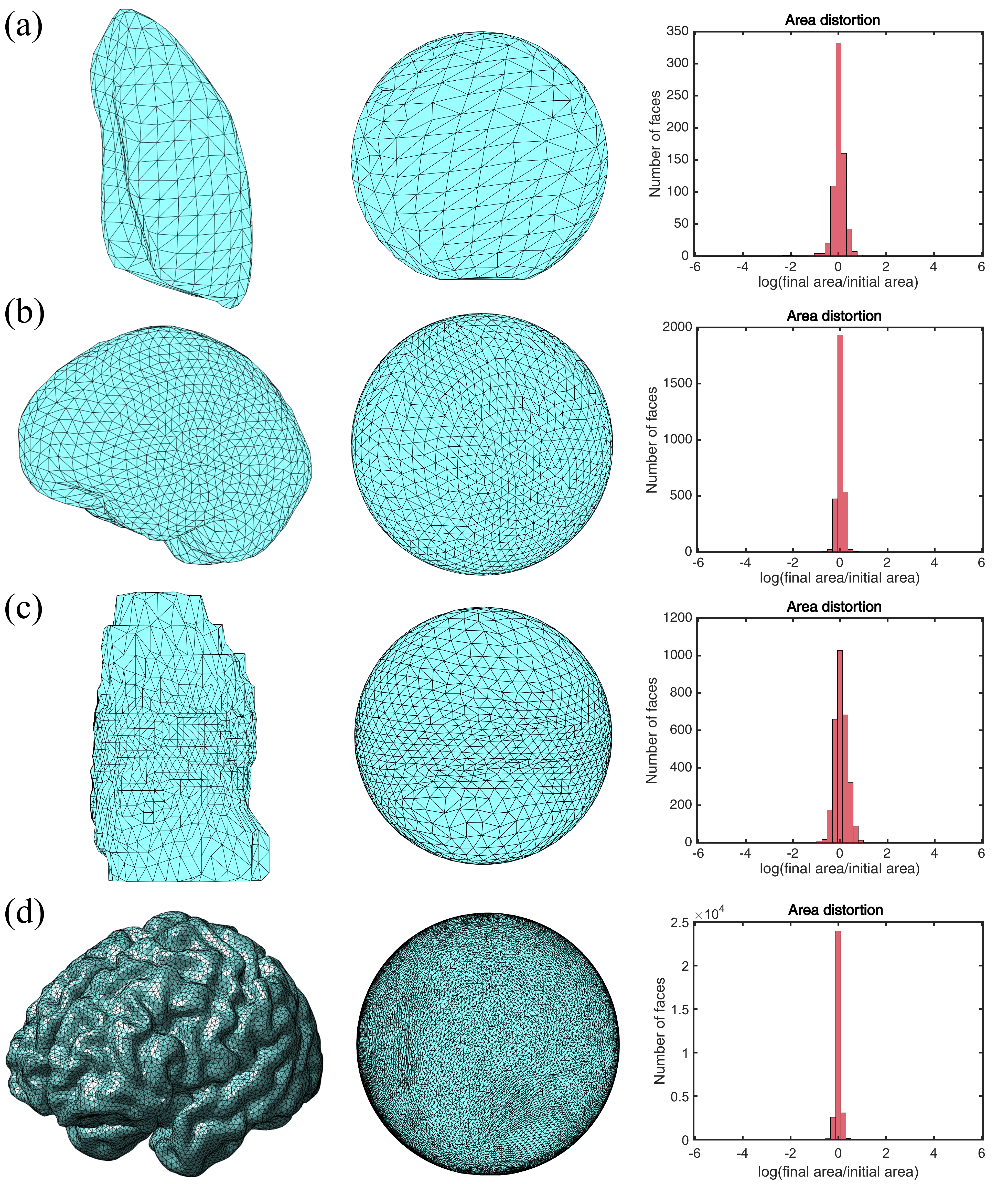}
    \caption{Adaptive area-preserving parameterization of genus-0 closed surfaces obtained by our proposed algorithm. (a)~A lung surface model from the CGTrader repository~\cite{CGTrader}. (b)~A human skull surface reconstructed from MRI images in the OASIS dataset~\cite{marcus2010open}. (c)~A human left ventricle surface reconstructed from MRI images in the ACDC dataset~\cite{bernard2018deep}. (d)~A human brain cortical surface from~\cite{lai2014folding}. For each example, the input surface, the adaptive area-preserving parameterization and the area distortion histogram are shown.} 
    \label{fig:mapping_result_closed}
\end{figure}

\begin{figure}[t!]
    \centering
    \includegraphics[width=0.48\textwidth]{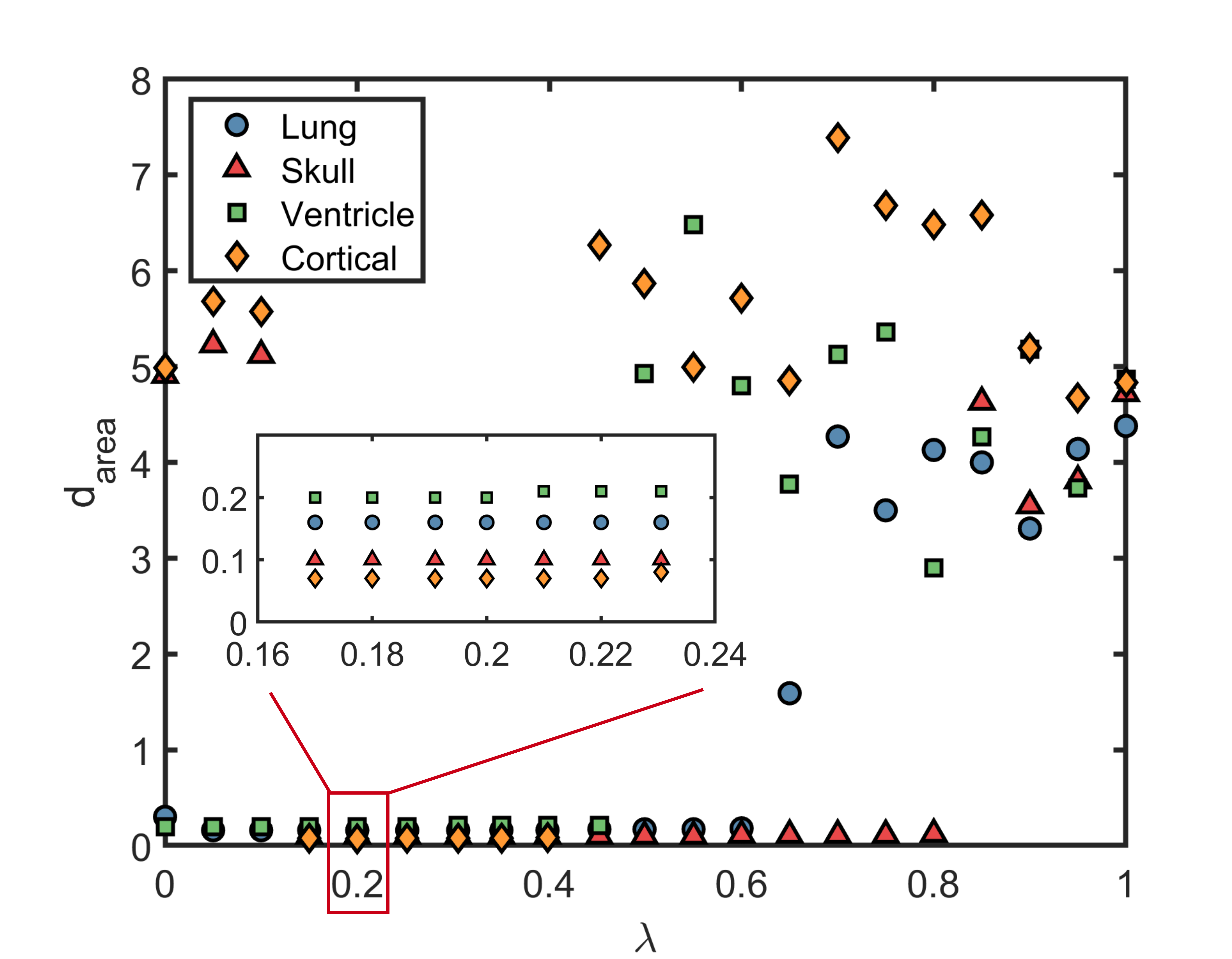}
    \caption{The effect of the initial flattening map parameter $\lambda$ on the adaptive area-preserving parameterization of genus-0 closed surfaces. The four surfaces in Fig.~\ref{fig:mapping_result_closed} are used in this experiment. Different values of $\lambda$  in $[0,1]$ are used for computing the adaptive parameterization, and the resulting area distortion $d_{\text{area}}$ is recorded for each $\lambda$ and each surface. The inset shows the results with more values of $\lambda$ considered at around $0.2$.} 
    \label{fig:lambda_experiments}
\end{figure}

\begin{table}[t!]
    \centering
    \begin{tabular}{c|c|c|C{22mm}}
    \multirow{ 2}{*}{Surface} & \multicolumn{3}{c}{mean$(|d_\text{area}|)$ / mean$(|d_\text{angle}|)$} \\ \cline{2-4}
         & Adaptive & Spherical & Hemispherical \\ \hline
         Fig.~\ref{fig:mapping_result_closed}(a) & 0.16 / 0.27 & 0.28 / 0.29 & 0.49 / 0.46\\ 
         Fig.~\ref{fig:mapping_result_closed}(b) & 0.10 / 0.13 & 0.10 / 0.14 & 0.66 / 0.31\\ 
         Fig.~\ref{fig:mapping_result_closed}(c) & 0.20 / 0.30 & 0.20 / 0.30 & 0.70 / 0.38\\ 
         Fig.~\ref{fig:mapping_result_closed}(d) & 0.07 / 0.22 & 0.10 / 0.23 & 0.79 / 0.40 
    \end{tabular}
    \caption{The performance of different methods for parameterizing genus-0 closed surfaces.}
    \label{tab:performance_closed}
\end{table}

\subsubsection{Parameterization of Closed Anatomical Surfaces}
For genus-0 closed anatomical surfaces, we first consider parameterizing a closed lung surface model (Fig.~\ref{fig:mapping_result_closed}(a)) freely available at the CGTrader repository~\cite{CGTrader} under the Royalty Free License and an intracranial volume (ICV) of closed human skull surface (Fig.~\ref{fig:mapping_result_closed}(b)) reconstructed from MRI images in the OASIS dataset~\cite{marcus2010open} using the FieldTrip toolbox in MATLAB (see~\cite{giri2021open} for more details of the reconstruction). Similar to the case of open surfaces, it can be observed from the distortion histograms that the adaptive parameterizations are highly area-preserving. Next, we reconstruct a human left ventricle surface using MRI images from the automated cardiac diagnosis challenge (ACDC)~\cite{bernard2018deep} and parameterize it using our algorithm (see Fig.~\ref{fig:mapping_result_closed}(c)). While the surface is relatively coarse and unsmooth, our method is capable of parameterizing it onto a smooth spherical cap region with very low area distortion. Finally, we consider a highly convoluted human brain cortical surface from~\cite{lai2014folding} (see Fig.~\ref{fig:mapping_result_closed}(d)). From the parameterization result and the area distortion histogram, it can be observed that our method works well even for surfaces with complicated geometry.

For comparison, we consider the spherical area-preserving parameterization (i.e. with $Z^* \approx -1$) and hemispherical area-preserving parameterization (i.e. with $Z^* = 0$). From Table~\ref{tab:performance_closed}, it can again be observed that our proposed method outperforms spherical and hemispherical parameterizations and achieves a significant improvement in the geometric distortion for handling surfaces with different geometry, which can be attributed to the flexibility of the adaptive domain.

\begin{figure*}[t!]
    \centering
    \includegraphics[width=0.9\textwidth]{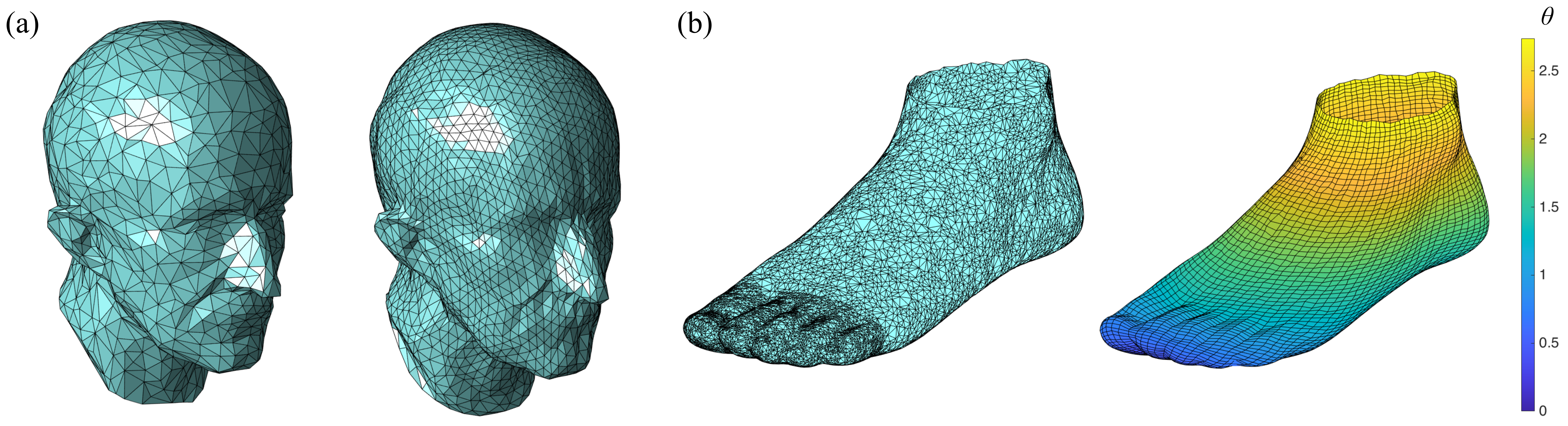}
    \caption{Surface remeshing via adaptive area-preserving parameterization. (a) By parameterizing a human head model with irregular triangulations onto an optimal spherical cap and generating a new regular mesh on the spherical cap, we can significantly improve the mesh quality of the surface. (b) We remesh a human foot model using the latitude and longitude lines defined on the optimal spherical cap. Here the color indicates the corresponding value of $\theta$.}
    \label{fig:remeshing}
\end{figure*}

Recall that for closed surfaces, the initial flattening map involves a combination of the disk conformal map and the SEM map via a balancing factor $\lambda$, which is set to be 0.2. It is natural to ask how the choice of $\lambda$ would affect the parameterization result. Here we compute the parameterization using different values of $\lambda$ and assess the area distortion $d_{\text{area}}$. As shown in Fig.~\ref{fig:lambda_experiments}, a small $\lambda \leq 0.1$ or a large $\lambda \geq 0.5$ may sometimes lead to a relatively large distortion in the final result. To explain this phenomenon, note that in general conformal maps focus on the preservation of angles without controlling the area, and so a more conformal initial map will usually contain highly squeezed triangle elements, thereby making the correction of the area distortion in the subsequent OMT computation more computationally challenging. Also, an initial map closer to the SEM map will usually involve a larger distortion in angle and hence many triangles may be highly irregular and stretched. In the computation of the power diagram and the subsequent OMT map, such irregular triangles may lead to large discretization errors and hence affect the numerical performance of the algorithm. By contrast, it can be observed that at around $\lambda = 0.2$, the result is robust to the value of $\lambda$. This experiment demonstrates the importance of the initial flattening map for the closed surface case and the robustness of the proposed method.

\subsection{Anatomical Surface Remeshing}
The proposed adaptive parameterization methods can be applied to surface remeshing for improving the quality of anatomical surface meshes. More specifically, to improve the mesh quality of any given anatomical surface $\mathcal{S}$, we can first compute the adaptive area-preserving parameterization $f:\mathcal{S} \to \mathbb{S}^2_{Z \geq Z^*}$ to map it onto an optimal spherical cap domain. We can then generate a regular triangle mesh on the spherical cap and map the new mesh back to the anatomical surface using the inverse mapping $f^{-1}$. As shown in Fig.~\ref{fig:remeshing}(a), the mesh quality of the input surface is significantly improved. More specifically, note that the adaptive parameterization-based remeshing method is advantageous in two aspects. First, as the parameterization is area-preserving and the triangle elements of the spherical cap mesh are highly uniform, the triangle elements of the resulting remeshed surface will also be highly uniform. Second, as the remeshed surface is generated using the inverse mapping $f^{-1}$, it is expected that the distance between the remeshed surface and the original surface is very small. 

To quantify the above properties, we first define the \emph{face area deviation} of the remeshed surface as
\begin{equation}
d_{\text{face}} =\mean_T \left|\text{Area}(T) - \text{mean face area}\right|,
\end{equation}
where $T$ is a triangular face in the remeshed surface. Note that $d_{\text{face}} = 0$ if and only if all triangle elements in the remeshed surface have the same face area. As for the distance between the remeshed surface and the original surface, we define the \emph{surface distance} as
\begin{equation}
d_{\text{surface}} = \mean_v \|v - \tilde{v}\|_2,
\end{equation}
where $v$ is a vertex in the remeshed surface and $\tilde{v}$ is the projection of $v$ onto the original surface. A small $d_{\text{surface}}$ indicates that the remeshed surface resembles the shape of the original surface well.

In Table~\ref{tab:remeshing}, we compare our parameterization-based remeshing approach with several other meshing methods available in the open-source 3D mesh processing software MeshLab~\cite{cignoni2008meshlab}. It can be observed that both the face area deviation $d_{\text{face}}$ and the surface distance $d_{\text{surface}}$ achieved by our method are lower than those by the other methods by over 50\% on average. This demonstrates that the effectiveness of our method for anatomical surface remeshing.

\begin{table}[t]
\centering
\begin{tabular}{c|c|c} 
Method & $d_{\text{face}}$ & $d_{\text{surface}}$ \\ \hline
Our proposed method & 0.0009 & 0.0034\\
Uniform Mesh Resampling~\cite{cignoni2008meshlab} & 0.0019 & 0.0088\\
Robust Implicit MLS~\cite{oztireli2009feature} & 0.0020 & 0.0072\\
Screened Poisson~\cite{kazhdan2013screened} & 0.0016 & 0.0085
\end{tabular}
\caption{The performance of our adaptive parameterization-based remeshing method and other methods. For a fair comparison, the target number of triangles in the remeshed surface is set to be around 5000 for all methods. For each method, the face area deviation $d_{\text{face}}$ and the surface distance $d_{\text{surface}}$ of the resulting remeshed surface are recorded.}
\label{tab:remeshing}
\end{table}

With the aid of the adaptive parameterization, it is also possible to remesh a surface using the latitude and longitude lines defined on the spherical cap. More specifically, since the spherical cap is a subdomain of the unit sphere, we can divide the spherical cap into regions using lines with the same latitude or longitude, which naturally induce a mesh grid on the original surface. This allows us to systematically divide the surface into different regions for further analysis (see Fig.~\ref{fig:remeshing}(b) for an example).

\begin{figure*}[t]
    \centering
    \includegraphics[width=\textwidth]{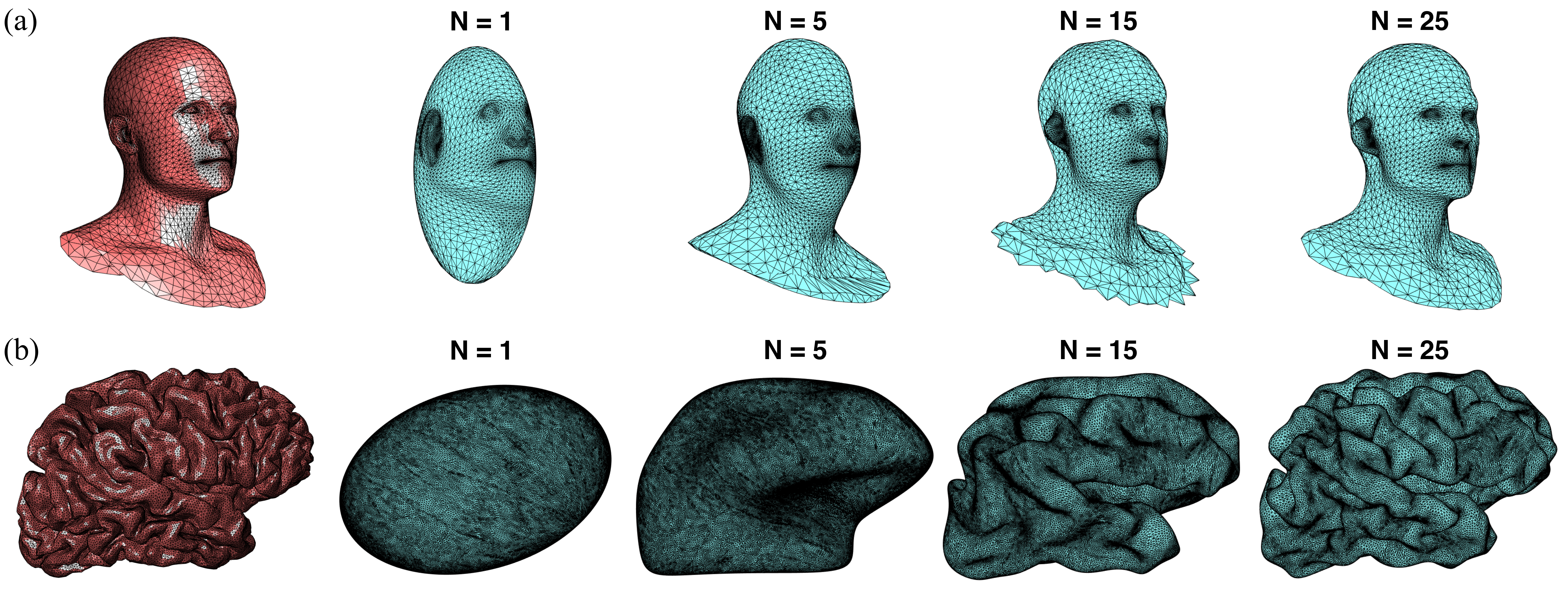}
    \caption{Anatomical shape description using the proposed adaptive parameterization and the AH basis functions. (a) A human face from the CGTrader repository~\cite{CGTrader} and the AH reconstructions with different maximum order $N$. (b) A highly convoluted brain cortical surface from the OASIS dataset~\cite{marcus2010open} and the AH reconstructions with different maximum order $N$.}
    \label{fig:ah_result}
\end{figure*}

\begin{figure*}[t]
    \centering
    \includegraphics[width=\textwidth]{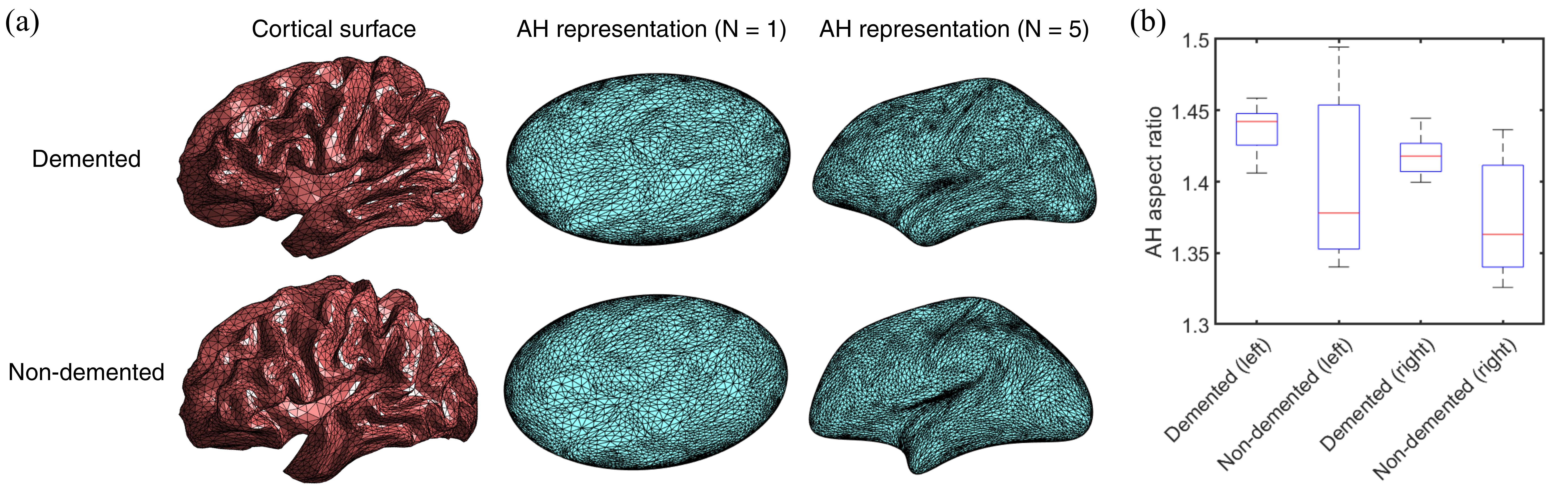}
    \caption{Shape analysis of demented cortical surfaces using the proposed adaptive parameterization and AH method. (a) Examples of demented and non-demented cortical surfaces from the OASIS dataset~\cite{marcus2010open}, with their low-order AH representations obtained from the proposed parameterization algorithm. (b) The aspect ratio of the low-order AH representation ($N=1$) for the left and right cortical surfaces for the 50 demented and non-demented subjects.}
    \label{fig:dementia_cortical}
\end{figure*}

\subsection{Anatomical Shape Description Using AH}
By combining the proposed adaptive parameterization and the AH basis functions, we can easily achieve a multilevel representation of any given anatomical surface. Fig.~\ref{fig:ah_result} shows two example anatomical surfaces and the AH reconstruction results with different maximum order $N$. It may be observed that even for $N = 1$, i.e. $(1+1)^2 = 4$ AH basis functions, the AH reconstructions are capable of capturing the overall geometry of the object surfaces. As $N$ increases, more details such as the ears of the human face and the gyri and sulci of the cortical surface can be captured.

\subsection{Shape analysis of demented and non-demented cortical surfaces}
It is natural to ask whether the proposed adaptive parameterization and AH method can be utilized for comparing different anatomical shapes. Here we consider 50 demented and non-demented subjects from the OASIS dataset~\cite{marcus2010open}. For each subject, we reconstruct the left and right cortical surfaces from the respective human head MRI scans in the dataset. We then apply our adaptive parameterization algorithm and obtain the low-order AH representations for each of them (see Fig.~\ref{fig:dementia_cortical}(a)). While it is difficult to compare the demented and non-demented cortical surfaces directly, one can see that the simplified, low-order AH representations of them are visually different. For a more quantitative comparison, note that the AH representation with $N=1$ gives an ellipsoidal geometry and hence can be used for quantifying the overall shape of the surface. To achieve this, we first approximate the AH representation using a matrix equation $\mathbf{y}_i \approx A\mathbf{x}_i$, where $\mathbf{y}_i$ are the coordinates of the AH representation, $A$ is a $3\times 3$ transformation matrix and $\mathbf{x}_i$ are the coordinates of the unit sphere with $\|\mathbf{x}\|_2 = 1$. More specifically, we solve the least-square problem 
\begin{equation}
\argmin_A \sum_{i=1}^n \|\mathbf{y}_i - A\mathbf{x}_i\|^2
\end{equation}
to get the optimal transformation matrix $A$. Then, we compute the singular value decomposition of the matrix $A$ and obtain the largest and smallest singular values $\sigma_1$ and $\sigma_3$. We can then assess the aspect ratio of the AH representation as $\frac{\sigma_1}{\sigma_3}$. As shown in the box plot in Fig.~\ref{fig:dementia_cortical}(b), the AH aspect ratios for the 50 demented and non-demented subjects are highly different. More specifically, the AH aspect ratios for both the left and right cortical surfaces of the demented subjects are higher than those of the non-demented ones on average and are also more concentrated. We further apply the two-sample $t$-test and found that the difference between the AH aspect ratio for the demented and the non-demented cortical surfaces is statistically significant for both the left brains ($p = 5.1 \times 10^{-5}$) and the right brains ($p =1.4\times 10^{-6}$). This demonstrates the clinical utility of the proposed adaptive parameterization and AH method for anatomical shape description and analysis.

\section{Conclusion}\label{sect:conclusion}
In this work, we have proposed two novel methods for parameterizing simply-connected open and closed anatomical surfaces. Unlike most prior methods, our methods treat the shape of the parameter domain as a variable in finding an optimal parameterization, resulting in an adaptive area-preserving parameterization onto an automatically determined spherical cap region on the unit sphere. Experimental results demonstrate the effectiveness of the proposed parameterization methods in comparison to the existing methods for both open and closed anatomical surfaces, including human scalp, tooth, face, lung, skull, ventricle and brain cortical surfaces. As our methods are area-preserving and also with minimal conformal distortion, they are advantageous for many biomedical applications including anatomical surface remeshing, shape description and shape analysis. In particular, the surface description of the object surfaces can be effectively achieved using a novel combination of the adaptive parameterization and AH. In the future, we plan to extend our method for parameterizing onto other adaptive domains such as a flexible ellipsoidal shape to handle more complicated geometries with singular points and different shape indexes~\cite{koenderink1990solid}. We also plan to apply the proposed parameterization methods and the AH basis functions for detecting shape anomaly in other anatomical datasets~\cite{nitzken20113d,nitzken20113d2}, thereby aiding disease prognosis and diagnosis.

\bibliography{adaptivebib}
\bibliographystyle{IEEEtran}

\end{document}